\documentclass[12pt]{article}
\makeatletter
\@addtoreset{equation}{section}
\makeatother

\newcommand{\be}{\begin{equation}}
\newcommand{\ee}{\end{equation}}
\newcommand{\bea}{\begin{eqnarray}}
\newcommand{\eea}{\end{eqnarray}}

\def\dt{\delta}
\def\om{\omega}
\def\bb{\bar{\beta}}

\def\G{\Gamma}
\def\Gt{\tilde{\Gamma}}
\def\tet{\tilde{\theta}}
\def\pst{\tilde{\psi}}
\def\gt{\tilde{G}}

\def\Ab{\bar{A}}
\def\P{\varphi}
\def\Pt{\tilde{\varphi}}
\renewcommand{\t}[1]{\tilde{#1}}

\begin{document}

\begin{titlepage}

\begin{center}
\vskip 2.5 cm
{\Large \bf Non-Abelian Duality in the String Effective Action}
\vskip 1 cm
{\large A. Bossard\footnote{e-mail:\ \tt bossard@celfi.phys.univ-tours.fr} 
and N. Mohammedi\footnote{e-mail:\ \tt nouri@celfi.phys.univ-tours.fr}}\\
\vskip 1cm
{\em Laboratoire de Math\'ematiques et Physique Th\'eorique\footnote{CNRS
UPRES-A 6083}\\
Universit\'e {}Fran\c{c}ois Rabelais \\
{}Facult\'e des Sciences et Techniques \\
Parc de Grandmont \\
{}F-37200 Tours - {}France.} 
\end{center}
\vskip 1.5 cm

\begin{abstract} 
\noindent
We study the symmetry of the one-loop effective action of bosonic string
theory under
non-Abelian T-duality transformations. 
It is shown that the original Lagrangian and its dual are proportional.
This result implies that the corresponding reduced low energy effective
actions are equivalent
and leads to a functional relation between the Weyl anomaly coefficients 
of the original and dual two-dimensional non-linear sigma models.
{}Finally, we apply this formalism to some simple examples. 

\end{abstract}

\end{titlepage}

       \section{Introduction}

Nowadays, duality and its different aspects are major tools  in the 
study of string theory.
One of these, target space duality
or  T-duality  connects seemingly different string background fields
\cite{buscher}.
It relies on the presence of isometries in the metric, the antisymmetric
tensor and the dilaton field.
The procedure to obtain the dual action corresponding to
a given non-linear sigma model is well known \cite{rocek}. It consists 
in gauging these isometries together with imposing, by means of a Lagrange
multiplier, 
that the gauge field strength vanishes. The elimination of the gauge
fields 
leads to a non-linear sigma model having the same global 
form as the original one but with different background fields.  
This means that a string propagating in a spacetime endowed with a
specific 
background metric, for instance,  presents the same dynamics when
moving in another spacetime having a completely different geometry.  
The T-duality transformations are said to be
Abelian \cite{buscher} or non-Abelian \cite{quevedo,aagl} according to the
nature of the Lie algebra
formed by the generators of the isometries. 
All of these considerations are, however, at the level of the classical
theory. 

One of the important questions, therefore,  is whether
the duality transformations remain symmetries  of the quantum theory.
A first step towards finding an answer to this difficult problem resides
in 
comparing the renormalization properties of the original sigma model and 
its dual.  One is led to considering the beta functions of the
two theories.  These should be studied regardless of whether 
the sigma model backgrounds provide consistent backgrounds in which 
the string can propagate (that is when the beta functions vanishes). 
Most of the investigations in this subject deal with Abelian T-duality. 
A relation between the beta functions of the original theory and their
counterparts
in the dual model has been given in \cite{haag}. This relation has been
reached by requiring
that the duality procedure commutes with the renormalization flow. In
other words,
one demands that $\left[{\rm T}\,,\,{{\rm d}\over {\rm d}\mu}\right]=0$,
where
T is the operator implementing the duality transformations and $\mu$ is
the scale of the 
renormalization group. At the one-loop level, the relation between the
beta functions
has been shown to hold for any sigma model possessing Abelian Killing
vectors \cite{haag}. 
At two loops \cite{bfhp,haagols,kaloper1,HaagOlsSchiap,Haagensen} and
three loops \cite{jack}, 
the same relation holds, provided that one 
modifies the duality transformations (used at the one-loop level)
order by order in perturbation theory. 

On the other hand, the situation regarding non-Abelian duality
transformations is still far from 
being completely understood. It was shown in \cite{tyurin}, using
supersymmetric arguments
that, if the original non-linear sigma model (with non-Abelian isometries)
is conformally
invariant at the one-loop level, then so is its non-Abelian dual.
This proof relies on the  
Curci-Paffuti \cite{curci} equations which relate different 
Weyl anomaly coefficients (the $\bar \beta$-functions).
Therefore not all of the relations between the $\bar \beta$-functions of
the two theories 
(the original and its dual) have been found in \cite{tyurin}.
Other explicit examples, where both
the original backgrounds and their corresponding non-Abelian duals are
consistent    
string backgrounds, have also been treated in \cite{giveon,ex1,ex2}. 
It is only recently that the general relation,
valid for both Abelian and non-Abelian T-duality, between 
the Weyl anomaly coefficients of the original theory and those
corresponding to its dual has been
given \cite{bfmps}. This relation is found through a path integral
procedure and for any 
(not necessarily conformally invariant) non-linear sigma model.
The only requirement is that 
the adjoint representation of the isometry Lie algebra is traceless
\cite{tyurin,giveon,faab,faab2}. 
This relation has been shown to hold at the one-loop level  for some
particular examples which are,
in general,  not conformally invariant \cite{bfmps,bbfmps,balog}.

The aim of this paper is to check the validity of the relation between the
Weyl anomaly
coefficients for a general sigma model and its non-Abelian dual. Clearly,
the case 
of the Abelian T-duality is also included here (simply by setting the
structure
constants of the isometry Lie algebra to zero).  
One of the nice properties of the Weyl anomaly coefficients of a
non-linear sigma model
is that they can be seen as linear combinations of the equations of motion
coming from a target space action known as the string theory effective
action \cite{callan}.
Our strategy is to examine the behaviour of this effective action
under non-Abelian T-duality. This allows us to find the precise relation
between the 
$\bar \beta$-functions of the original theory and those of its non-Abelian
dual.
It turns out that this relation is precisely the one found in \cite{bfmps}
through the
partition function of the sigma model. This result provides another check
on the intimate connection between the two-dimensional theory
(where the duality transformations are found) and the target space string
effective action.
It is worth mentioning that the study of the string effective action under
Abelian T-duality
has revealed a rich structure: the duality transformations found through
the sigma model are 
only a particular case of a family of symmetries that form an $O\left(d,d,
R\right)$ group \cite{odd}.

The paper is organized as follows:  
in the next section we recall the procedure
used to obtain the dual sigma model 
and give the non-Abelian T-duality transformations.
We show that, by choosing a suitable reparametrization of the different
backgrounds, 
a simplified version of these transformations can be found. 
{}Furthermore, this reparametrization is essential in carrying out the
explicit 
computation of the different terms in the original string effective action
and in its dual. 
To get an idea of what kind 
of relationship one might expect between the effective action and its
dual,
we start, in section 3, by supposing that the functional relation of
\cite{bfmps} between 
the Weyl anomaly coefficients does indeed hold. This implies that the
Lagrangian 
of the dual string effective action must be proportional to the Lagrangian
of the original action.   
This proportionality is then checked  
by a direct calculation of the dual Lagrangian, 
followed by a comparison with the initial Lagrangian.
In section 5 we show how the formalism works 
for two simple examples. One of these is based 
on a Bianchi V type metric \cite{bianchi5} and is known to be inequivalent
to its dual at the 
quantum level (the adjoint representation of the isometry Lie algebra is
not
traceless). Nevertheless, the two reduced string effective actions (the
original and its dual) 
are equivalent up to an overall factor.   
{}Finally, in section 7, we present our main conclusions and 
give possible developments of the present work. 
An appendix lists the essential formulas of the paper, 
especially those used in the explicit calculations of section 4.

\section{The duality transformations}

In this section our main goal is to obtain the T-duality transformations
of
the background fields. 
Let us consider a general non-linear sigma model of the form 

\begin{eqnarray}
S[x,y] & = & \int d\sigma ^{+}d\sigma ^{-}\{\partial _{+}x^{\mu }Q_{\mu
\nu }(x)\partial _{-}x^{\nu }
+\partial _{+}x^{\mu }Q_{\mu a}(x)e^a_i(y)\partial_-y^i \nonumber \\
 & + & 
e^a_i(y)\partial_+y^iQ_{a\mu }(x)\partial _{-}x^{\mu }
+e^a_i(y)\partial_+y^iQ_{ab}(x)e^b_j(y)\partial_-y^j\label{original} 
\\
 & - & \frac{1}{4}R^{\left( 2\right) }\varphi (x)\}\,\,.\nonumber 
\end{eqnarray}
The scalar field \( \varphi (x) \) is the dilaton and \( R^{(2)} \) is the
two 
dimentional curvature of the worldsheet. 
As usual \( \partial _{\pm } \) denotes differentiation
with respect to $\sigma ^{\pm }$. Here $e^a_i$ are vielbeins defined by 
$(g^{-1}\partial _{\pm }g)=T_ae^a_i\partial_{\pm}y^i$
where the \( T_{a} \) are the generators of a Lie algebra $\cal{G}$
and $g$ is an element of the corresponding Lie group. They verify
\begin{equation}
\label{fee}
\partial _{i}e_{j}^{a}-\partial
_{j}e_{i}^{a}=-f_{bc}^{a}e_{i}^{b}e_{j}^{c},
\end{equation}
where $f^a_{bc}$ are the structure constants of $\cal{G}$. 
The inverses of these vielbeins are denoted $E^i_a$ 
and satisfy $E_{a}^{m}\partial _{m}E^{k}_{b}-E_{b}^{m}\partial
_{m}E^{k}_{a}=f^{c}_{ab}E_{c}^{k}$.
 
It is clear that the action (\ref{original}) is invariant under the
transformation 
$g\rightarrow ug$ where $u$ is a constant element of the Lie group.
As it has been done in \cite{quevedo,aagl,giveon}, the dual model of
the action (\ref{original}) is found by gauging this isometry
transformation.
Introducing gauge fields $A^a_+$ and $A^a_-$ and choosing the gauge $g=1$, 
yields the first order action 
\begin{eqnarray}
S_1[x,\lambda ,A] & = & \int d\sigma ^{+}d\sigma ^{-}
\{\partial _{+}x^{\mu }Q_{\mu \nu }\partial _{-}x^{\nu }+A_{+}^{a}Q_{a\mu
}\partial _{-}x^{\mu }
\nonumber \\
 &+& \partial _{+}x^{\mu }Q_{\mu a}A_{-}^{a}+A^{a}_{+}Q_{ab}A^{b}_{-}
+\lambda _{c}(\partial _{+}A^{c}_{-}-\partial
_{-}A^{c}_{+}+f_{ab}^{c}A_{+}^{a}A_{-}^{b})\nonumber \\
 &-& \frac{1}{4}R^{\left( 2\right) }\varphi \}\,\,.
\label{base}
\end{eqnarray} 
The equations of motion of the Lagrange multiplier $\lambda_c$ force the
fields strength ${}F_{+-}^c=
(\partial _{+}A^{c}_{-}-\partial _{-}A^{c}_{+}+
f_{ab}^{c}A_{+}^{a}A_{-}^{b})$ to vanish. The solution to these
constraints
is $A^a_{\pm}=e^a_i\partial_{\pm}y^i$. Putting this solution into
(\ref{base}) 
yields the original action (\ref{original}). On the other hand,
integrating over the gauge fields, instead, leads to the dual action
\begin{eqnarray}
\tilde{S}[x,\lambda ] & = & \int d\sigma ^{+}d\sigma^{-}
\{\partial _{+}x^{\mu }Q_{\mu \nu }\partial _{-}x^{\nu }-
(\partial _{+}x^{\mu }Q_{\mu a}-\partial _{+}\lambda _{a})N^{ab}
(Q_{b\nu }\partial _{-}x^{\nu }+\partial _{-}\lambda _{b})\nonumber \\
 &  & -\frac{1}{4}R^{\left( 2\right) }\Pt \}\,\,,
\label{dual} 
\end{eqnarray}
where $N^{ab}$ is the inverse matrix of 
\begin{equation}
	M_{ab}=Q_{ab}+\lambda_{c}f_{ab}^{c}\,\,.
\label{Mab}
\end{equation}
Notice that one can find a kind of inverse
operation which permits one to recover
$S_1[x,\lambda ,A]$ from the dual action $\tilde{S}[x,\lambda]$.
This can be seen by considering the following action
\begin{eqnarray}
S_2[x,\lambda ,C,\chi ] & = & \int d\sigma ^{+}d\sigma ^{-}
\{\partial _{+}x^{\mu }Q_{\mu \nu }\partial _{-}x^{\nu }
-(\partial _{+}x^{\mu }Q_{\mu a}-C_{+a})N^{ab}
(Q_{b\nu }\partial _{-}x^{\nu }+C_{-b})\nonumber \\
 &  & -\chi _{+}^{a}(C_{-a}-\partial _{-}\lambda _{a})
+(C_{+a}-\partial _{+}\lambda _{a})\chi _{-}^{a}-\frac{1}{4}R^{\left(
2\right) }\varphi \}\,\,.
\label{sigma-back} 
\end{eqnarray}
The integration over the new Lagrange mutipliers $\chi_{\pm}^a$ results
in the dual action $\tilde{S}[x,\lambda]$. 
However, if one keeps $\chi_{\pm}^a$ and integrates over 
$C_{+a}$ and $C_{-a}$, then one arrives 
to the action $S_1[x,\lambda,\chi]$ where
$\chi _{\pm}^a$ play the role of the gauge fields $A_{\pm}^a$. 
It is important to point out that this inverse operation 
does not rely on the existence of any isometries. 
This suggests that duality transformations 
are not necessarily related to the presence of 
isometries as is the case in Poisson-Lie duality \cite{poisson}.

In order to compare the dual action with the original one, 
let us make the field redefinition $\lambda _{a}=y^{i}\eta _{ia}$. 
For convenience, we have used the same symbol $y^i$ to denote
the original field and its dual.  
We choose $\eta_{ia}$
to be invertible and symmetric with determinant equal to one.
To obtain the duality transformations on the backgrounds,
the original sigma model is written as
\begin{equation}
S[x,\lambda ] = \int d\sigma ^{+}d\sigma^{-}
\{\partial_+z^M(G_{MN}+B_{MN})\partial_-z^N
 -\frac{1}{4}R^{\left( 2\right) }\varphi \}\,\,, 
\label{GBmodel}
\end{equation}
where $z^N=(x^\mu,y^i)$. The dual sigma model is of 
the same form and its dual background fields
are denoted $\tilde{G}_{MN}$, $\tilde{B}_{MN}$ and $\Pt$.
The duality transformations are then given by
\begin{equation}
\label{Trelations}
\begin{array}{ccl}
\tilde{G}_{\mu \nu } & = & 
G_{\mu \nu }-\bar{S}^{ab}E^{i}_{a}E^{j}_{b}(G_{\mu i}G_{\nu j}
-B_{\mu i}B_{\nu j})+\bar{A}^{ab}E^{i}_{a}E^{j}_{b}(G_{\mu i}B_{\nu
j}-B_{\mu i}G_{\nu j})\\
\tilde{B}_{\mu \nu } & = & B_{\mu \nu
}-\bar{A}^{ab}E^{i}_{a}E^{j}_{b}(G_{\mu i}G_{\nu j}
-B_{\mu i}B_{\nu j})+\bar{S}^{ab}E^{i}_{a}E^{j}_{b}(G_{\mu i}B_{\nu
j}-B_{\mu i}G_{\nu j})\\
\tilde{G}_{\mu i} & = & -(G_{\mu k}E_{a}^{k}\bar{A}^{ab}+B_{\mu
k}E_{a}^{k}\bar{S}^{ab})\eta _{bi}\\
\tilde{B}_{\mu i} & = & -(G_{\mu k}E_{a}^{k}\bar{S}^{ab}+B_{\mu
k}E_{a}^{k}\bar{A}^{ab})\eta _{bi}\\
\tilde{G}_{ij} & = & \eta _{ia}\bar{S}^{ab}\eta _{bj}\\
\tilde{B}_{ij} & = & \eta _{ia}\bar{A}^{ab}\eta _{bj}\\
\tilde{\varphi } & = & \varphi -\frac{1}{2}\ln \left( \det M_{ab}\right)
\,\,. 
\end{array}
\end{equation}
Here we have written $N^{ab}=\bar{S}^{ab}+\bar{A}^{ab}$ where
$\bar{S}^{ab}$ is symmetric and $\bar{A}^{ab}$ is antisymmetric.
Notice that the transformation of the dilaton does
not come from the classical procedure described above. 
In fact it is a purely quantum effect. 
We refer to \cite{buscher,odd,dilaton} for details 
regarding the origin of this term. 
Independently we shall see that this dilaton shift is 
indeed the one needed
for the equality between the reduced string effective actions 
of the original and the dual backgrounds.

{}For later use, let us write $Q_{ab}=S_{ab}(x)+v_{ab}(x)$ where
$S_{ab}$ is symmetric and $v_{ab}$ is antisymmetric. This leads
to the decomposition $M_{ab}=S_{ab}+A_{ab}$ where
\begin{equation}
A_{ab}(x,y) = v_{ab}(x)+y^{i}\eta _{ic}f^{c}_{ab}\,\,.
\end{equation}
Using the decompositions of $N^{ab}$ and $M_{ab}$ together 
with the relation $N^{ab}M_{bc}=\delta^a_c$, we arrive at
the crucial relations    
\begin{eqnarray}
S_{ab}\bar{S}^{bc}+A_{ab}\bar{A}^{bc} & = & \delta^c_a\nonumber \\
S_{ab}\bar{A}^{bc}+A_{ab}\bar{S}^{bc} & = & 0\,\,.
\label{ssbar} 
\end{eqnarray}
These will be used extensively in the following sections.

The T-duality transformations, as given in (\ref{Trelations}), 
are quite complicated, especially
for $G_{\mu \nu }$ and $B_{\mu \nu }$. 
It is therefore convenient
to look for  a suitable decomposition of the backgrounds, 
in which the duality transformations are simpler. 
As in the Abelian case \cite{haag,kaloper1}, 
we choose a decomposition \`a la Kaluza-Klein.
A reparametrization of the metric, consistent with the original 
sigma model, is given by
\begin{equation}
\label{metG}
G_{MN}=\left( \begin{array}{ll}
g_{\mu \nu }(x)+h_{ij}V_{\mu }^{i}V_{\nu }^{j} \,\,\,& V^{i}_{\mu
}h_{ij}\\
V^{i}_{\mu }h_{ij} & h_{ij}
\end{array}\right) 
\end{equation}
with 
\begin{equation}
\label{h_ij}
h_{ij}(x,y)=e_{i}^{a}(y)S_{ab}(x)e^{b}_{j}(y)\,\,\,,\,\,\,
V_{\mu}^i(x,y)=E^i_{a}(y)X_{\mu}^a(x)\,\,.  
\end{equation}
Similarly, the antisymmetric tensor is decomposed as
\begin{equation}
\label{B_MN}
B_{MN}=\left( \begin{array}{ll}
b_{\mu \nu }(x)-\frac{1}{2}(V^{k}_{\mu }b_{\nu k}-V^{k}_{\nu }b_{\mu k}) 
\,\,\,& b_{\mu i}\\
-b_{\mu i} & b_{ij}
\end{array}\right) 
\end{equation}
where
\begin{equation}
\label{b_ij}
b_{ij}(x,y)=e^{a}_{i}(y)v_{ab}(x)e^{b}_{j}(y)\,\,\,,\,\,\,
b_{\mu i}(x,y)=e_i^a(y)W_{\mu a}(x)\,\,. 
\end{equation}
In this decomposition the dual backgrounds are also of the form
\begin{eqnarray}
\tilde{G}_{MN}&=&\left( \begin{array}{ll}
\t{g}_{\mu \nu }+\tilde{h}_{ij}\tilde{V}^{i}_{\mu }\tilde{V}^{j}_{\nu } 
\,\,\,& \tilde{V}^{i}_{\mu }\tilde{h}_{ij}\\
\tilde{V}^{i}_{\mu }\tilde{h}_{ij} & \tilde{h}_{ij}
\end{array}\right)\nonumber\\
\tilde{B}_{MN}&=& \left( \begin{array}{ll}
\tilde{b}_{\mu\nu}
-\frac{1}{2}(\tilde{V}^{k}_\mu \tilde{b}_{\nu k}
-\tilde{V}^{k}_{\nu }\tilde{b}_{\mu k}) 
\,\,\,& \tilde{b}_{\mu i}\\
-\tilde{b}_{\mu i} & \tilde{b}_{ij}
\end{array}\right)\,\,.
\end{eqnarray}
The transformations (\ref{Trelations}) become
\begin{eqnarray}
\tilde{g}_{\mu \nu } & = & g_{\mu \nu }\nonumber \\
\tilde{b}_{\mu \nu } & = & b_{\mu \nu }\nonumber \\
\tilde{V}_{\mu }^{i} & = & (V_{\mu }^{l}e_{l}^{a}A_{ab}
-b_{\mu l}E_{b}^{l})\eta ^{bi}\nonumber \\
\tilde{b}_{\mu i} & = & -(V^{l}_{\mu }e_{l}^{a}S_{ab}\bar{S}^{bc}
+b_{\mu l}E_{b}^{l}\bar{A}^{bc})\eta _{ci}\nonumber \\
\tilde{h}_{ij} & = & \eta _{ia}\bar{S}^{ab}\eta _{bj}\nonumber \\
\tilde{b}_{ij} & = & \eta _{ia}\bar{A}^{ab}\eta _{bj}\nonumber \\
\tilde{\varphi } & = & \varphi -\frac{1}{2}\ln \left( \det
M_{ab}\right)\,\,. 
\label{T-rel-simple} 
\end{eqnarray}
In this manner, \( g_{\mu \nu } \) and \( b_{\mu \nu } \) are invariants 
and the duality transformations act only on 
$ V_{\mu }^{i}$, $b_{\mu i}$, $h_{ij}$, $b_{ij}$ and the dilaton
$\varphi$.
As it is well known, 
the Kaluza-Klein parametrization has other important advantages. 
{}Firstly, it enables one to calculate the inverse metric 
straightforwardly and, consequently, the scalar curvature.
In our case, the inverse of $G_{MN}$ is 
\begin{equation}
\label{invmetG}
G^{MN}=\left( \begin{array}{ll}
g^{\mu \nu } \,\,\,& -V^{\mu i} \\
-V^{\mu i} & h^{ij}+V^{i}_{\mu }V^{\mu j}
\end{array}\right) 
\end{equation}
where Greek indices are raised and lowered with the metric $g_{\mu \nu}$
and $h^{ij}$ is the inverse of $h_{ij}$.  
Secondly, the determinant of the metric $G_{MN}$, 
is simply given by
\begin{equation}
\label{det}
\det G_{MN}=\det g_{\mu \nu }\det h_{ij}\,\,.
\end{equation}
These last two properties will be useful in the next sections.

\section{Weyl anomaly coefficients and string effective action}

We will deal in what follows with the low energy effective action 
of string theory at one-loop. We will examine its behaviour under
non-Abelian
T-duality transformations. This effective action is given by
\begin{equation}
\label{eff_act}
\Gamma ^{(1)}[G,B,\varphi ]=\int d^Dz \sqrt{G}\exp (-2\varphi )
\{R(G)+4\partial _{M}\varphi \partial ^{N}\varphi
-\frac{1}{12}H_{MNP}H^{MNP}+\Lambda\}\,\,.
\end{equation}
In this expression, $R$ is the scalar curvature of the metric $G_{MN}$ and
$H_{MNP}$, defined by 
$H_{MNP}=\partial _{M}B_{NP}+\partial _{N}B_{PM}+\partial _{P}B_{MN}$,
is the torsion of the antisymmetric field $B_{MN}$.
We have also chosen to include a cosmological constant $\Lambda$
which should be set to zero for critical strings. 
{}For this task, it is more convenient 
to use the Kaluza-Klein decomposition of the previous section.
The coordinates are split as $z^M=(x^\mu,y^i)$  where $M=1,...,D$,  
$\mu=1,...,d$ and $i=d+1,...,D$.

Before getting into the details of the calculations, 
we will list some useful relations.
Using equations (\ref{ssbar}) we can show, in matrix notation, that
\begin{equation}
M\bar{S}=SN^{t}
\end{equation}
where $N^{t}$ stands for the transpose of the matrix $N^{ab}$.
As a direct consequence we have
\begin{equation}
\label{det_M}
\det M=\sqrt{\frac{\det S}{\det \bar{S}}}\,\,.
\end{equation}
The Kaluza-Klein form of the metric $G_{MN}$ leads to
\begin{equation}
\det G_{MN}=(\det e_{i}^{c})^{2}\det g_{\mu \nu }\det S_{ab}\,\,.
\end{equation}
Similarly, the determinant of the dual metric $\tilde{G}_{MN}$ 
is given by
\begin{equation}
\det \tilde{G}_{MN}=\det g_{\mu \nu }\det \bar{S}_{ab}
\end{equation}
due to the fact that we have chosen $\det \eta_{ia} =1$.
These last three equations combined with the definition of the dual
dilaton result in
\begin{equation}
\label{Rel_1}
\sqrt{\det \tilde{G}}\exp (-2\tilde{\varphi })=
\frac{1}{\det e}\sqrt{\det G}\exp (-2\varphi )\,\,.
\end{equation}
Note that the two sides of this equation are both independant of
the coordinate $y^i$. This last expression confirms the good choice for
the dilaton transformation, as it implies proportionality between
the integration measures. 

{}Furthermore, the effective action 
$\Gamma [G,B,\varphi]=\int d^Dz \sqrt{G} \exp(-2\varphi) L$
can be expressed, up to an integration by parts, 
as
\begin{eqnarray}
\label{R_H_phi}
L[G,B,\P]&=&R(G)-\frac{1}{12}H_{MNP}H^{MNP}
-4\partial_{M} \varphi \partial^{M} \varphi
+4\nabla_{M} \partial^{M} \varphi +\Lambda 
\nonumber\\
&=&G^{MN}\bar{\beta }_{MN}^{G}
-4\bar{\beta }^{\varphi }\,\,,
\end{eqnarray}
where the Weyl anomaly coefficients of the sigma model (\ref{GBmodel}) 
are given by
\begin{eqnarray}
\bar{\beta }^{G}_{MN} & = & R_{MN}+2\nabla _{M}\partial _{N}\varphi 
-\frac{1}{4}H_{MPQ}H^{\; \; \; PQ}_{N} \nonumber \\
\bar{\beta }^{B}_{MN} & = & -\frac{1}{2}\nabla ^{P}H_{MNP}+H_{MNP}\partial
^{P}\varphi \nonumber \\
\bar{\beta }^{\varphi } & = & -\frac{1}{2}\nabla ^{2}\varphi 
+\partial _{P}\varphi \partial ^{P}\varphi
-\frac{1}{24}H_{MNP}H^{MNP}-\frac{\Lambda}{4}\,\,.
\end{eqnarray}

Let us turn now to a crucial relationship between the Weyl anomaly
coefficients
($\bb^{G},\bb^{B},\bb^{\P}$)
of the original sigma model and the Weyl anomaly coefficients 
($\bb^{\tilde{G}},\bb^{\tilde{B}},\bb^{\Pt}$)
of its corresponding dual.
At the one-loop level and for Abelian T-duality, these relations 
have been given in \cite{haag,haagols}.
The non-Abelian counterparts of these relations
have been found in \cite{bfmps}.
These relations have been derived using path integral arguments
with the restriction $f^{a}_{ab}=0$ on the structure constants
of the isometry Lie algebra \cite{tyurin,faab}.
In order to express these relations in a compact form, we will
use the generic notation $\omega_{A}=(G_{MN},B_{MN},\P)$
with $\omega_1=G_{MN},\dots$.
Similarly the dual backgrounds are denoted 
$\tilde{\omega}_{A}=(\tilde{G}_{MN},\tilde{B}_{MN},\Pt)$. 
In this notation we have
\begin{equation}
\bb^{\t{\omega}_A}=
\frac{\delta \t{\omega}_A}{\delta \omega_B }\bb^{\omega_B}
\label{rel_beta}
\end{equation}
These relations have been shown to hold for a few simple sigma models.
However, no proof is known for a general sigma model.
The aim of this paper is to show the validity of equations
(\ref{rel_beta})
for a sigma model of the type considered in the previous section.
{}For this, we will use the string effective action.

As a starting point, let us suppose that equations (\ref{rel_beta}) 
hold and let us see what the consequences are for the effective actions
$\Gamma[\omega]$ and $\tilde{\Gamma}[\tilde{\omega}]$.
With the help of (\ref{Rel_1})
we have 
\begin{eqnarray}
\gt^{MN}\bb^{\gt}_{MN}
-4\bb^{\Pt} 
& = & 
\{ \gt^{MN}\frac{\dt \gt_{MN}}{\dt \om_A }\bb^{\omega_A}
-4\frac{\dt \Pt}{\dt\om_A }\}
\nonumber \\
 & = & 
2\bb^{\omega_A}\frac{\dt}{\dt \om_A}
\ln (\sqrt{\det \gt}\exp (-2\Pt))
\nonumber \\
 & = & 
\bb^{\omega_A}\frac{\dt}{\dt \om_A}
\{\ln (\det G)-2\varphi -\ln (\det e)\}
\nonumber \\
 & = & 
G^{MN}\bb^{G}_{MN}-4\bb^{\P }
\end{eqnarray}
where we have used $\frac{\delta e}{\delta \omega}=0$.
Thus we have
\begin{eqnarray}
\Gamma[\omega]&=&\int d^dx d^ny 
\det(e)\det(g)\det(S)\exp(-2\P)L[\omega]\nonumber\\
\Gamma[\tilde{\omega}]&=&\int d^dx d^ny \det(g)\det(S)\exp(-2\P)L[\omega]
\end{eqnarray} 
where $d+n=D$.
Therefore, the reduced effective actions (the original and its dual)
would be proportional if one were able to perform the integration
over the coordinates $y^i$.
This demands that $L[\omega]$ does not depend on $y^i$.
It turns out that this is indeed the case as explained in what follows.

{}From the form of the metric of the previous section 
it can be shown, after straightforward but tedious calculations, that
\begin{eqnarray}
R(G) & = & R(g)-\frac{1}{4}S_{ab}X^{a}_{\mu \nu }X^{b\mu \nu }
-\frac{1}{4}S^{ab}D_{\mu }S_{ab}S^{cd}D^{\mu }S_{cd}
+\frac{3}{4}S^{ab}D_{\mu }S_{bc}S^{cd}D^{\mu }S_{da} \nonumber \\
 &&-S^{ab}D_{\mu}D^{\mu}S_{ab}
-\frac{1}{4}f^{a}_{ce}f^{b}_{df}S_{ab}S^{cd}S^{ef}
-\frac{1}{2}f^{a}_{cb}f^{b}_{da}S^{cd}-f^{a}_{ca}f^{b}_{db}S^{cd}
\label{R_x} 
\end{eqnarray}
where 
\begin{equation}
X^{a}_{\mu \nu }\equiv \partial _{\mu }X^{a}_{\nu }
-\partial _{\nu }X^{a}_{\mu }+f^{a}_{bc}X_{\mu }^{b}X_{\nu }^{c}
\end{equation}
and \( D_{\mu } \) is the gauge covariant derivative defined by
\begin{equation}
D_{\mu }S_{ab}=\partial _{\mu }S_{ab}
-f^{c}_{da}X_{\mu }^{d}S_{cb}-f^{c}_{db}X_{\mu }^{d}S_{ca}\,\,.
\end{equation}
The expression (\ref{R_x})
shows that $R(G)$ is a function of 
$g_{\mu \nu }(x)$, $S_{ab}(x)$
and $X^{a}_{\mu \nu }(x)$. Hence it is independent of $y^i$.

Similarly, using the decomposition of $G_{MN}$ in (\ref{metG}),
it can be shown that
\begin{equation}
H^2=H_{MNP}H^{MNP}=(h_{\mu \nu \rho }h^{\mu \nu \rho })+3(h_{\mu \nu
i}h^{\mu \nu i})
+3(h_{\mu ij}h^{\mu ij})+(h_{ijk}h^{ijk})
\label{hh}
\end{equation}
where we raise Greek indices with \( g^{\mu \nu } \) and Latin
indices with $h^{ij}(x,y)=E_{a}^{i}(y)S^{ab}(x)E^{j}_{b}(y)$ 
with $S^{ab}$ the inverse of $S_{ab}$.
Here the components of $h_{MNP}$ are defined by
\begin{eqnarray}
h_{ijk} & = & H_{ijk}\nonumber \\
h_{\mu ij} & = & H_{\mu ij}-V^{k}_{\mu }H_{kij}\nonumber \\
h_{\mu \nu i} & = & H_{\mu \nu i}-\{V^{k}_{\mu }H_{k\nu i}
-(\mu \leftrightarrow \nu )\}+V^{k}_{\mu }V^{l}_{\nu }H_{kli}\nonumber \\
h_{\mu \nu \rho } & = & H_{\mu \nu \rho }-\{V^{k}_{\mu }H_{k\nu \rho }
+\mathbf{c.p.}\}+\{V^{k}_{\mu }V^{l}_{\nu }H_{kl\rho }
+\mathbf{c.p.}\}-V^{k}_{\mu }V^{l}_{\nu }V^{m}_{\rho }H_{klm}
\label{h_MNP} 
\end{eqnarray}
where $\mathbf{c.p.}$ stands for cyclic permutations.
Using the decomposition (\ref{B_MN}) of the antisymmetric tensor $B_{MN}$, 
one finds that
\begin{eqnarray}
h_{ijk} & = & 
e^a_ie^b_je^c_k\{v_{ad}f^d_{bc}+{\bf c.p.}\}
\nonumber\\
h_{\mu ij} & = & 
e^a_ie^b_j\{\partial_\mu v_{ab}+W_{\mu d}f^d_{ab}-X^c_\mu
(v_{ad}f^d_{bc}+{\bf c.p.})\}
\nonumber \\
h_{\mu \nu i} & = & 
e^a_i\{[\partial_\mu W_{\nu a}-X^b_\mu ( \partial_\nu v_{ab}
+W_{\nu d}f^d_{ab})-(\mu \leftrightarrow \nu)]
+X_\mu^b X_\nu^c (v_{bd}f^d_{ca}+{\bf c.p.})\} 
\nonumber \\
h_{\mu \nu \rho } & = & 
\partial_\mu(b_{\nu\rho}
-\frac{1}{2}X_\nu^aW_{\rho a}+\frac{1}{2}X_\rho^a W_{\nu a})
-X_\mu^a(\partial_\nu W_{\rho a}-\partial_\rho W_{\nu a})\nonumber\\
&+&X_\mu^aX_\nu^b(\partial_\rho v_{ab}+W_{\rho d}f^d_{ab})
-X_\mu^aX_\nu^bX_\rho^c(v_{ad}f^d_{bc})
+{\bf c.p.}
\end{eqnarray}
We see that when performing the different contractions
in (\ref{hh}), the vielbeins $e^a_i(y)$ simplify with their inverses
$E_a^i(y)$ coming from $h^{ij}$. Hence, the dependence on $y^i$ disappears
from $H^2$. 

{}Finally, the dilaton contribution is given by
\begin{equation}
4(\nabla_M\partial^M\P-\partial_M\P\partial^M\P)=
4g^{\mu\nu}\left[\nabla_\mu \partial_\nu \P 
-\partial_\mu\P \partial_\nu\P
+\partial_\mu \P \left(\frac{1}{2} S^{ab} \partial_\nu S_{ab} 
+ X_\nu^b f^a_{ba}\right)\right]\,\,.
\end{equation}
This expression is explicitly independent of $y^i$.
We conclude therefore that 
$L[\omega]=G^{MN}\bar{\beta }^{G}_{MN}-4\bar{\beta }^{\varphi }$
is independant of $y^i$.

\section{Invariance of the string effective action}

In this section, we will prove the relation (\ref{rel_beta}) between 
the Weyl anomaly coefficients of the original sigma model
and those of its dual.
Our strategy consists in obtaining this relation 
from the invariance of the string effective action
under non-Abelian T-duality transformations.
We begin our study from the relationship between the 
Weyl anomaly coefficients and the string effective action.
As it is well known, these are related by
\begin{equation}
\bb^{\omega_A}=M_{AB}[\om]
\frac{\delta S[\omega]}{\delta \omega_B}
\label{bbm1}
\end{equation}
where the matrix $M$ takes the form 
\begin{equation}
M_{AB}[\om]=-\frac{1}{\sqrt{G}\exp(-2\P)}
\left(\begin{array}{lll}
\frac{1}{2}(G_{MP}G_{NQ}+G_{MQ}G_{NP})\,\,&0\,\,&\frac{1}{4}G_{PQ}\\
0&\frac{1}{2}(G_{MP}G_{NQ}-G_{MQ}G_{NP})&0\\
\frac{1}{4}G_{MN}&0&\frac{1}{16}(D-2)
\end{array}\right)
\label{Mmat}
\end{equation}
Of course the same relations hold for the dual background
$\tilde{\omega}$ and we have
\begin{equation}
\bb^{\tilde{\omega}_A}=M_{AB}[\t{\om}]
\frac{\delta S[\tilde{\omega}]}{\delta \tilde{\omega}_B}\,\,.	
\label{bbm2}
\end{equation}
Suppose that we have 
\begin{equation}
S[\tilde{\omega}]=A[\omega]S[\omega]\,\,\,\,{\rm where}\,\,\,\,
S[\omega]=\sqrt{G}\exp(-2\P)L[\omega]\,\,\,\,,
\,\,\,\,S[\tilde{\omega}]=\sqrt{\tilde{G}}\exp(-2\Pt)L[\tilde{\omega}]
\label{prop}
\end{equation}
then, using a chain rule and the expressions (\ref{bbm1}, \ref{bbm2}),
we deduce that 
\begin{equation}
\bb^{\tilde{\omega}_A}=M_{AB}[\t{\om}]
\frac{\delta A[\omega]}{\delta \tilde{\omega}_B}S[\omega]
+A[\omega]M_{AB}[\t{\om}]
\frac{\delta \omega_C}{\delta
\tilde{\omega}_B}M^{CD}[\t{\om}]\bb^{\omega_D}
\end{equation}
where $M^{CD}[\t{\om}]$ denotes the inverse matrix of $M_{AB}[\t{\om}]$.
This last expression is comparable to (\ref{rel_beta}) if one has
\begin{equation}
\frac{\delta A}{\delta \tilde{\omega}}=0\,\,\,,\,\,\,
A[\omega]M_{AB}[\t{\om}]
\frac{\delta \omega_C}{\delta \tilde{\omega}_B}M^{CD}[\t{\om}]=
\frac{\delta \tilde{\omega}_A}{\delta \omega_D}\,\,.
\label{Eq2}
\end{equation} 
Our aim is therefore to show that equations (\ref{prop}) and (\ref{Eq2})
are indeed satisfied.
This will be our task in this section.

We start by proving that equation (\ref{prop}) is fulfilled.
One cannot check this at the level of the reduced expressions as given 
in the previous section. 
This is due to the fact that in computing those reduced terms,
we have thrown away some expressions whose transformed counterparts
are non zero. {}For instance, we have, in the original effective action
$\Gamma[\omega]$,
terms such as $\partial_i \P(x)$, which obviously vanish.
However, the equivalent term in the dual effective action 
$\Gamma[\tilde{\omega}]$ is $\partial_i \Pt(x,y)$, 
which is different from zero.
This situation is to be contrasted with Abelian T-duality, where 
the original backgrounds and their duals are both independent of the 
coordinates $y^i$.
Hence the above problem is not encountered in the Abelian case. 
In order to avoid this potential source of errors, 
one is forced to work with
a string effective action where the dependence on the coordinates
$x^\mu$  and $y^i$ is not a priori specified. 
The particular coordinate dependence of the backgrounds
is made explicit only once the duality transformations are carried out.

Checking that $S[\tilde{\omega}]=A[\omega]S[\omega]$ is a tremendous task.
Here, we will give a highlight of the calculations 
and leave the details for the appendix.
Moreover, we will write down only the expressions corresponding
to the original effective action $\Gamma[\omega]$.
Those corresponding to the dual effective action $\Gamma[\tilde{\omega}]$
are simply obtained by replacing each tensor by a tilded one:
its T-duality transformed tensor as given in (\ref{T-rel-simple}).  
The Kaluza-Klein decomposition of the metric $G_{MN}$ 
yields the Ricci scalar\footnote{We have used the package Ricci, 
developped by J.M. Lee, to perform most of the calculations here.}
\begin{eqnarray}
R(G) & = & R(g)
\nonumber \\
 && +\{
h^{ik}h^{jl}\partial _{i}\partial _{j}h_{kl}
-h^{ij}h^{kl}\partial _{i}\partial _{j}h_{kl}
 \nonumber \\
 && -\frac{1}{2}h^{ir}h^{jl}h^{ks}\partial _{i}h_{jk}\partial _{l}h_{rs}
+\frac{3}{4}h^{il}h^{jr}h^{ks}\partial _{i}h_{jk}\partial _{l}h_{rs} 
\nonumber \\
 && -h^{ij}h^{kr}h^{ls}\partial _{i}h_{jk}\partial _{l}h_{rs}
-\frac{1}{4}h^{il}h^{jk}h^{rs}\partial _{i}h_{jk}\partial _{l}h_{rs}
+h^{ij}h^{kl}h^{rs}\partial _{i}h_{jk}\partial _{l}h_{rs}
\} 
\nonumber\\
 && +g^{\mu \nu }\{
(-h^{ij}\nabla _{\mu }\nabla _{\nu }h_{ij}
+\frac{3}{4}h^{ik}h^{jl}\nabla _{\mu }h_{ij}\nabla _{\nu }h_{kl}
-\frac{1}{4}h^{ij}h^{kl}\nabla _{\mu }h_{ij}\nabla _{\nu }h_{kl})
\nonumber  \\
 && +(2\partial _{i}\nabla _{\mu }V_{\nu }^{i}
+h^{ik}\partial _{i}V_{\nu }^{j}\nabla _{\mu }h_{jk}
+2h^{jk}V^{i}_{\mu }\nabla _{\nu }\partial _{i}h_{jk}
+h^{jk}\partial _{i}h_{jk}\nabla _{\mu }V^{i}_{\nu } \nonumber \\
 && +h^{jk}\partial _{i}V_{\mu }^{i}\nabla _{\nu }h_{jk}
-\frac{3}{2}h^{jl}h^{kr}V^{i}_{\mu }\nabla _{\nu }h_{lr}\partial
_{i}h_{jk}
+\frac{1}{2}h^{jk}h^{lr}V^{i}_{\mu }\nabla _{\nu }h_{lr}\partial
_{i}h_{jk}) \nonumber \\
 && +(-2V_{\mu }^{j}\partial _{i}\partial _{j}V_{\nu }^{j}
-\frac{1}{2}\partial _{i}V^{j}_{\mu }\partial _{j}V^{i}_{\nu }
-\partial _{i}V^{i}_{\mu }\partial _{j}V^{j}_{\nu }
-\frac{1}{2}h^{ik}h_{jl}\partial _{i}V^{j}_{\mu }\partial _{k}V_{\nu }^{l}
\nonumber \\
 && -h^{il}V^{k}_{\mu }\partial _{i}V^{j}_{\nu }\partial _{k}h_{jl}
-h^{kl}V^{i}_{\mu }\partial _{i}V^{j}_{\nu }\partial _{j}h_{kl}
-h^{kl}V^{j}_{\mu }\partial _{i}V^{i}_{\nu }\partial _{j}h_{kl} \nonumber
\\
 && -h^{kl}V^{i}_{\mu }V^{j}_{\nu }\partial _{i}\partial _{j}h_{kl}
-\frac{1}{2}h^{ir}h^{ks}V^{j}_{\mu }V^{l}_{\nu }\partial
_{i}h_{jk}\partial _{l}h_{rs}
+\frac{1}{2}h^{jl}h^{ks}V^{i}_{\mu }V^{r}_{\nu }\partial
_{i}h_{jk}\partial _{l}h_{rs} \nonumber \\
 && +\frac{3}{4}h^{jr}h^{ks}V^{i}_{\mu }V^{l}_{\nu }\partial
_{i}h_{jk}\partial _{l}h_{rs}
-\frac{1}{4}h^{jk}h^{rs}V^{i}_{\mu }V^{l}_{\nu }\partial
_{i}h_{jk}\partial _{l}h_{rs})\} \nonumber \\
 && -\frac{1}{4}g^{\mu \rho}g^{\nu \lambda}h_{ij}
(V^{i}_{\mu \nu}-{}F^{i}_{\mu \nu})
(V^{j}_{\rho \lambda}-{}F^{j}_{\rho \lambda})
\end{eqnarray}
with the definitions 
\begin{equation}
V^{i}_{\mu \nu} =  
\partial_{\mu } V^{i}_{\nu}
-\partial_{\nu} V^{i}_{\mu}
\,\,\,{\rm and}\,\,\,
{}F^{i}_{\mu \nu}  =  
V_{\mu}^{k} \partial_k V_{\nu}^{i}
-V_{\nu}^{k}\partial_k V_{\mu}^{i}\,\,.
\end{equation}
The terms have been assembled according to the number of factors of 
$g^{\mu\nu}$ and the number of factors of $V_\mu^i$. 
This separation of terms will serve as a guide in our calculation.
Since $g_{\mu\nu}$ is T-duality invariant then 
$R(g)$ is obviously invariant too.

Let us turn our attention now to the dilatonic contribution.
In Abelian duality it is convenient to reparametrize the dilaton
in a T-duality invariant manner \cite{odd}. 
This simplifies greatly the calculations.
Unfortunately this happens to be impossible here due
to the transformation properties of the internal metric $h_{ij}$. 
The best we can do is to reparametrize the dilaton as 
\begin{equation}
\label{dil1}
\varphi (x)=\psi (x,y)+\theta(x,y).
\end{equation}
where we define $\theta=\frac{1}{4}\ln \det h_{ij}$.
Under duality transformations, we have 
$\P \rightarrow \Pt=\P-\frac{1}{4}\ln\det S+\frac{1}{4}\ln\det \bar{S}$
and $\theta \rightarrow \tilde{\theta}=\frac{1}{4}\ln\det \bar{S}$.
We deduce therefore
the following transformation for $\psi$
\begin{equation}
\label{dil2}
\psi \rightarrow \tilde{\psi }(x)=\psi (x,y)+\frac{1}{2}\ln \det e(y)\,\,.
\end{equation}
This new scalar field $ \psi $  has
an interesting property: its dual depends only on $x$
(this can be seen by replacing, in the last expression, $\psi$ by 
$\P - \theta$). 
Using the decomposition (\ref{dil1}), the dialtonic contribution 
to the original effective action is 
\begin{eqnarray}
4G^{MN}(\nabla _{M}\partial _{N}\varphi -\partial _{M}\varphi \partial
_{N}\varphi ) 
& = & 
4[G^{\mu \nu }\partial _{\mu }\partial _{\nu }\psi 
-G^{\mu \nu }\partial _{\mu }\psi \partial _{\nu }\psi\nonumber\\&& 
+2G^{\mu i}\partial _{i}\partial _{\mu }\psi 
-G^{MN}\Gamma ^{\lambda }_{MN}\partial _{\lambda }\psi\nonumber\\&&  
-2G^{\mu i}\partial _{\mu }\psi \partial _{i}\psi 
-2G^{\mu M}\partial _{\mu }\psi \partial _{M}\theta\nonumber\\&& 
+G^{ij}\partial _{i}\partial _{j}\psi 
+G^{MN}\partial _{M}\partial _{N}\theta\nonumber\\&& 
-G^{MN}\Gamma ^{i}_{MN}\partial _{i}\psi 
-G^{MN}\Gamma ^{P}_{MN}\partial _{P}\theta\nonumber\\&& 
-G^{ij}\partial _{i}\psi \partial _{j}\psi 
-2G^{iM}\partial _{i}\psi \partial _{M}\theta\nonumber\\&& 
-G^{MN}\partial _{M}\theta\partial _{N}\theta]\,\,.
\label{dil3} 
\end{eqnarray}
The following expressions for the Christoffel symbols are useful 
\begin{eqnarray}
G^{MN}\Gamma ^{\lambda }_{MN}(G) & = & g^{\mu \nu }\Gamma ^{\lambda }_{\mu
\nu }(g)
+g^{\lambda \alpha }[(-\frac{1}{2}h^{ij}\nabla _{\alpha }h_{ij})
+(\partial _{i}V^{i}_{\alpha }+\frac{1}{2}V^{k}_{\alpha }h^{ij}\partial
_{k}h_{ij})]\label{gam1} \\
G^{MN}\Gamma ^{k}_{MN}(G) & = & (h^{ij}h^{kl}\partial _{i}h_{jl}
-\frac{1}{2}h^{ik}h^{jl}\partial _{i}h_{jl})+g^{\alpha \beta }(\nabla
_{\alpha }V^{k}_{\beta }
+\frac{1}{2}V^{k}_{\beta }h^{ij}\nabla _{\alpha }h_{ij})\nonumber \\
 &  & +g^{\alpha \beta }[-\partial _{i}(V^{i}_{\alpha }V_{\beta }^{k})
-\frac{1}{2}V^{i}_{\alpha }V^{k}_{\beta }h^{jl}\partial
_{i}h_{jl}]\label{gam2} 
\end{eqnarray}

The contribution coming from the torsion term, $H^2$, is determined next.
The expressions in (\ref{h_MNP}) together with the decomposition
(\ref{B_MN})
lead to   
\begin{eqnarray}
h_{ijk} & = & \partial _{i}b_{jk}+\mathbf{c}.\mathbf{p}.\\
h_{\mu ij} & = & (\partial _{\mu }b_{ij})+(-V^{k}_{\mu }\partial
_{k}b_{ij}
+\{\partial _{j}b_{\mu i}'+(\partial _{j}V^{k}_{\mu
})b_{ki}-(i\leftrightarrow j)\})\\
h_{\mu \nu i} & = & (b_{\mu \nu i}'+V_{\mu \nu }^{k}b_{ki})
+([V^{k}_{\mu }b_{i\nu k}'-(\mu \leftrightarrow \nu )]
-\frac{1}{2}\partial _{i}U_{\mu \nu }-{}F^{k}_{\mu \nu }b_{ki})\\
h_{\mu \nu \rho } & = & (\partial _{\rho }b_{\mu \nu })
+(-V_{\rho }^{k}\partial _{k}b_{\mu \nu })
+\frac{1}{2}(V^{k}_{\mu \rho }b_{\nu k}'+b_{\mu \rho k}'V_{\nu
}^{k})\nonumber \\
 &  & 
+\frac{1}{2}(F^{k}_{\rho \mu }b_{\nu k}'+V_{\rho }^{k}V_{\mu }^{l}b_{l\nu
k}')+{\bf c.p.}
\end{eqnarray}
where we have defined
\begin{eqnarray}
b_{\mu i}' & = & b_{\mu i}-V^{k}_{\mu }b_{ki}\nonumber\\
b_{\mu \rho i}' & = & \partial _{\mu }b_{\rho i}'-\partial _{\rho }b_{\mu
i}'
\nonumber\\
b_{i\rho j}' & = & \partial _{i}b_{\rho j}'-\partial _{j}b_{\rho i}'
\nonumber\\
U_{\mu \nu } & = & V^{k}_{\mu }b_{\nu k}'-V^{k}_{\nu }b_{\mu k}'\,\,.
\end{eqnarray}
The introduction of these new tensors is motivated by 
their simplified T-duality transformation rules, as seen below.
In terms of these tensors we have
\begin{eqnarray}
-\frac{1}{12}H_{MNP}H^{MNP}&=&	
\{+\frac{1}{2}h^{km}h^{il}h^{jn}\partial _{k}b_{ij}\partial _{l}b_{mn}
-\frac{1}{4}h^{kl}h^{im}h^{jn}\partial _{k}b_{ij}\partial _{l}b_{mn}\}
\nonumber\\
&&+ g^{\mu\nu}h^{im}h^{jn}
\{(
\partial_{\mu}b_{ij}\partial_{m}b_{\nu n}'
+
\partial_{\mu}b_{ij}\partial_{m}V^{k}_{\nu}b_{kn}
\nonumber\\
&&+\frac{1}{2}
\partial_{\mu}b_{ij}V^{k}_{\nu}\partial_{k}b_{mn})
\nonumber\\
&&+(-\frac{1}{4}V^{k}_{\mu }V^{l}_{\nu }
\partial_{k}b_{ij}\partial_{l}b_{mn}
-V^{l}_{\mu }
\partial _{l}b_{ij}\partial _{m}V^{k}_{\nu }b_{kn}
\nonumber\\
&&-V^{l}_{\mu }
\partial _{l}b_{mn}\partial _{i}b_{\nu j}'
-\frac{1}{2}
\partial _{m}V^{l}_{\mu }b_{ln}
[\partial_i V^{k}_{\nu }b_{kj}-(i\leftrightarrow j)]
\nonumber\\
&&-
\partial _{m}V^{k}_{\mu}b_{kn}
[\partial _{i}b_{\nu j}'-(i\leftrightarrow j)]
\nonumber\\
&&-\frac{1}{2}
\partial _{m}b_{\mu n}'
[\partial _{i}b_{\nu j}'-(i\leftrightarrow j)])\}
\nonumber\\
&&-\frac{1}{4}g^{\mu \rho}g^{\nu \lambda}h^{ij}
(h^{(1)}_{\mu \nu i}+h^{(2)}_{\mu \nu i})
(h^{(1)}_{\rho \lambda j}+h^{(2)}_{\rho \lambda j})
\nonumber\\
&& -\frac{1}{12}g^{\mu\alpha}g^{\nu\beta}g^{\rho\gamma}
h_{\mu\nu\rho}h_{\alpha\beta\gamma}
\label{hcarre}
\end{eqnarray}
where we have defined 
\begin{equation}
h^{(1)}_{\mu\nu i}=b_{\mu\nu i}'+V_{\mu\nu}^{k}b_{ki}
\,\,\,,\,\,\,
h^{(2)}_{\mu\nu i}=[V^{k}_{\mu }b_{i\nu k}'-(\mu \leftrightarrow \nu )]
-\frac{1}{2}\partial _{i}U_{\mu \nu }-{}F^{k}_{\mu \nu }b_{ki}\,\,.
\end{equation}
The terms of (\ref{hcarre}) have been also written according 
to the number of factors  of $g_{\mu\nu}$, $V_\mu^i$ and $b_{\mu i}$.  

Having listed the different contributions to $L[\omega]$ 
and $L[\tilde{\omega}]$, we are in a position to compare them.
The easiest part to start with is the dilatonic contribution involving
the field $\psi$ in both actions.
As a consequence of (\ref{dil2}), we have 
$\partial _{\mu }\tilde{\psi }=\partial_{\mu }\psi$.
Notice that $\partial_{\mu }\psi$ and $\partial _{\mu }\tilde{\psi}$ 
can only come from the dilatonic parts of the two actions.
Therefore, by identifying these terms in the two actions using
(\ref{dil3}), 
we must have 
\begin{eqnarray}
\label{dil4}
G^{\mu \nu }\partial _{\mu }\partial _{\nu }\psi 
-G^{\mu \nu }\partial _{\mu }\psi \partial _{\nu }\psi 
+2G^{\mu i}\partial _{i}\partial _{\mu }\psi 
-G^{MN}\Gamma ^{\lambda }_{MN}\partial _{\lambda }\psi 
-2G^{\mu i}\partial _{\mu }\psi \partial _{i}\psi 
-2G^{\mu M}\partial _{\mu }\psi \partial _{M}\theta
&=&\nonumber\\
\gt^{\mu \nu }\partial _{\mu }\partial _{\nu }\pst 
-\gt^{\mu \nu }\partial _{\mu }\pst \partial _{\nu }\pst 
+2\gt^{\mu i}\partial _{i}\partial _{\mu }\pst 
-\gt^{MN}\Gt ^{\lambda }_{MN}\partial _{\lambda }\pst 
-2\gt^{\mu i}\partial _{\mu }\pst \partial _{i}\pst 
-2\gt^{\mu M}\partial _{\mu }\pst \partial _{M}\tet
\end{eqnarray}
The first two terms on each side of this equality are equal
since $G^{\mu \nu }=\tilde{G}^{\mu \nu }=g^{\mu \nu }$.
The third term on each side vanishes due to the particular 
dependence on $x^\mu$ and $y^i$ of $\psi$ and $\tilde{\psi}$.
An explicit calculation of the remaining terms leads to
\begin{equation}
\frac{1}{2}V^{k}_{\sigma }h^{ij}\partial _{k}h_{ij}+\partial
_{k}V^{k}_{\sigma }=
\partial _{k}\tilde{V}^{k}_{\sigma }\,\,.
\end{equation}
Using the explicit form of $\tilde{V}^{k}_{\sigma }$, as given in
(\ref{T-rel-simple}), we find that this last equation is satisfied
only when 
\begin{equation}
V^{k}_{\sigma }e_{k}^{b}f^{a}_{ab}=0
\label{Vef}
\end{equation}
This equation holds if the isometry Lie algebra is such that
$f^{a}_{ab}=0$, 
as is the case for semi-simple Lie algebras. 
This observation agrees with the conclusions
reached in \cite{tyurin, giveon,faab, faab2}
for the quantum validity of non-Abelian T-duality transformations. 
However, the terms containing $f^a_{ab}$ contain always  
$V^{k}_{\sigma}$ at the same time.
Therefore, for a  block-diagonal metric with $V^{k}_{\sigma}=0$
the above equation is also satisfied.  
This particular case will be discussed in
an example based on a Bianchi-type metric in the next section.

The other particular terms to consider in comparing $\Gamma[\omega]$
and $\Gamma[\tilde{\omega}]$ are the contributions involving
three factors of $g^{\mu\nu}$.
These are found in the expressions of $H^2$ and $\tilde{H}^2$.
According to (\ref{hcarre}),
we must have 
\begin{equation}
g^{\mu\alpha}g^{\nu\beta}g^{\rho\gamma}h_{\mu\nu\rho}h_{\alpha\beta\gamma}
=
\tilde{g}^{\mu\alpha}\tilde{g}^{\nu\beta}\tilde{g}^{\rho\gamma}
\tilde{h}_{\mu\nu\rho}\tilde{h}_{\alpha\beta\gamma}\,\,.
\end{equation}
Since $g^{\mu\nu}=\tilde{g}^{\mu\nu}$, we must have
\begin{equation}
h_{\mu\nu\rho}=\tilde{h}_{\mu\nu\rho}\,\,.
\end{equation}   
We have checked, using the list of transformations 
given in the appendix, that this is indeed the case.

The comparison of the remaining terms
in $L[\om]$ and $L[\t{\om}]$ is organised 
according to the following pattern:
If we look at the transformations (\ref{T-rel-simple}), 
we note two important properties. 
{}First, since the reduced metric $g^{\mu \nu }$ is invariant, 
the terms in the two Lagrangians with the same
number of factors of $g^{\mu \nu }$ 
must be equal.
Moreover, $\t{V}_{\mu }^{i}$ and $\t{b}_{\mu i}$ 
contain one $V_{\mu}^i$ and one $b_{\mu i}$ factors. 
Therefore, the terms in $L[\t{\om}]$, consisting of $n$ factors of
$\t{V}_{\mu }^{i}$ and $m$ factors of $\t{b}_{\mu i}$
can only be compared with terms in $L[\om]$ containing $n'$ factors of
$\t{V}_{\mu }^{i}$ and $m'$ factors of $\t{b}_{\mu i}$,
such that $n'+m'=n+m$.  
Using this criterion as a guiding principle,
we have explicitly checked that $L[\t{\om}]=L[\om]$.
We have listed in an appendix the transformations of
all the terms. 

Having shown that $S[\t{\om}]=A[\omega]S[\om]$, with $A[\omega]=1/\det e$,
we can now turn our attention to proving the relations in (\ref{Eq2}).
It is clear that 
$\frac{\dt A[\om]}{\dt \t{\om}_B}
=\frac{\dt A[\om]}{\dt \om_A}\frac{\dt \om_A}{\dt \t{\om}_B}=0$.
In order to deal with the second equation in (\ref{Eq2}),
let us start by showing that all the matrices involved there 
are well defined. The right hand side requires the computation
of $\frac{\dt \t{\om}_A}{\dt \om_B}$.
{}For this, one needs to compute the variations 
$\dt S_{ab}$ and $\dt A_{ab}$. 
Using the key relations (\ref{ssbar}), we find
\bea
\frac{\dt \bar{S}^{ab}}{\dt {S}_{cd}}&=& 
-\frac{1}{2}[\bar{S}^{ac}\bar{S}^{db}
+\bar{A}^{ac}\bar{A}^{db}+(c \leftrightarrow d)]\nonumber\\
\frac{\dt \bar{S}^{ab}}{\dt {A}_{cd}}&=& 
-\frac{1}{2}[\bar{S}^{ac}\bar{A}^{db}
+\bar{A}^{ac}\bar{S}^{db}-(c \leftrightarrow d)]\nonumber\\
\frac{\dt \bar{A}^{ab}}{\dt {S}_{cd}}&=& 
-\frac{1}{2}[\bar{S}^{ac}\bar{A}^{db}
+\bar{A}^{ac}\bar{S}^{db}+(c \leftrightarrow d)]\nonumber\\
\frac{\dt \bar{A}^{ab}}{\dt {A}_{cd}}&=&
-\frac{1}{2}[\bar{S}^{ac}\bar{S}^{db}
+\bar{A}^{ac}\bar{A}^{db}-(c \leftrightarrow d)]\,\,.
\eea  
These expressions, together with $\frac{\dt e^a_i}{\dt \om_A}=0$,
and the duality transformations (\ref{Trelations}) 
yield the different components of $\frac{\dt \t{\om}_A}{\dt \om_B}$.
These are set forth in the appendix. 

The matrix $M^{CD}[\omega]$, inverse of $M_{AB}[\omega]$,
is computed from the variation of the action, where
$\frac{\delta S[\omega]}{\delta \omega_C}=M^{CD}[\om]\bb^{\omega_D}$ and
is given by
\begin{eqnarray*}
&&M^{CD}[\om]=
-\sqrt{G}\exp(-2\P)\times\nonumber\\
&&\left(\begin{array}{lcr}
\frac{1}{2}(G^{MP}G^{NQ}+G^{MQ}G^{NP}-G^{MN}G^{PQ})\,\,&0\,\,&2G^{PQ}\\
0&\frac{1}{2}(G^{MP}G^{NQ}-G^{MQ}G^{NP})&0\\
2G^{MN}&0&8
\end{array}\right)
\end{eqnarray*}
On the other hand, the computation of 
$\frac{\delta \omega_C}{\delta \tilde{\omega}_B}$
requires the inversion of the duality transformations.
The original string backgrounds can expressed in terms
of the dual ones as follows
\begin{equation}
\label{Tinvrelations}
\begin{array}{ccl}
G_{\mu \nu } & = & \t{G}_{\mu \nu }
-S_{ab}\eta^{ai}\eta^{bj}(\t{G}_{\mu i}\t{G}_{\nu j}-\t{B}_{\mu
i}\t{B}_{\nu j})
+A_{ab}\eta^{ai}\eta^{bj}(\t{G}_{\mu i}\t{B}_{\nu j}-\t{B}_{\mu
i}\t{G}_{\nu j})\nonumber\\
B_{\mu \nu } & = & \t{B}_{\mu \nu }
-A_{ab}\eta^{ai}\eta^{bj}(\t{G}_{\mu i}\t{G}_{\nu j}-\t{B}_{\mu
i}\t{B}_{\nu j})
+S_{ab}\eta^{ai}\eta^{bj}(\t{G}_{\mu i}\t{B}_{\nu j}-\t{B}_{\mu
i}\t{G}_{\nu j})\nonumber\\
G_{\mu i} & = & -(\t{G}_{\mu k}\eta^{ka}A_{ab}+\t{B}_{\mu
k}\eta^{ka}S^{ab})e^b_i\nonumber\\
B_{\mu i} & = & -(\t{G}_{\mu k}\eta^{ka}S_{ab}+\t{B}_{\mu
k}\eta^{ka}A^{ab})e^b_i\nonumber\\
G_{ij} & = & e_i^aS_{ab}e^b_j\\
B_{ij} & = & e_i^aA_{ab}e^b_j-y^k\eta_{kc}f^c_{ab}e^a_ie^b_j\\
\P & = & \Pt -\frac{1}{2}\ln \det (\bar{S}^{ab}+\bar{A}^{ab})\,\,.
\end{array}
\end{equation}
The different functionnal derivatives are then calculated in
a similar way to those of $\frac{\dt \t{\om}_A}{\dt \om_B}$
(recall that $\frac{\dt}{\dt
\t{\om}_B}(y^k\eta_{kc}f^c_{ab}e^a_ie^b_j)=0$).

Since the inverses of the matrices $M_{AB}[\om]$ and
 $\frac{\dt \om_A}{\dt \t{\om}_B}$ exist,
the second relation in (\ref{Eq2}) can be put in the form
 \be
A[\omega]M_{AB}[\t{\om}]=
\frac{\dt \t{\om}_A}{\dt \om_C}
M_{CD}[\om]
\frac{\dt \t{\om}_D}{\dt \om_B}
\label{Mt2}
\ee 
Using the explicit form of $M_{AB}[\om]$ in (\ref{Mt2}) and
the expressions of $\frac{\dt \t{\om}_A}{\dt \om_B}$ given in the
appendix,
we have checked that the above equation is indeed satisfied.
Let us just give an example of how this works.
The ($\t{G}_{ij},\Pt$) component of the left hand side equation
(\ref{Mt2})
is 
\be
A[\om]M_{\t{G}_{ij},\Pt} = -\frac{1}{4\det e
\sqrt{\t{G}}\exp(-2\Pt)}\t{G}_{ij}
\ee 
while the left hand side is written as 
\be
\frac{\dt \t{G}_{ij}}{\dt G_{mn}}M_{G_{mn}, \P}\frac{\dt \Pt}{\dt \P}
+  \frac{\dt \t{G}_{ij}}{\dt G_{mn}}M_{G_{mn}, G_{pq}}\frac{\dt \Pt}{\dt
G_{pq}}
+  \frac{\dt \t{G}_{ij}}{\dt B_{mn}}M_{B_{mn}, B_{pq}}\frac{\dt \Pt}{\dt
B_{pq}}
\ee
where 
\bea
M_{G_{mn}, \P}&=&-\frac{1}{4 \sqrt{G}\exp(-2\P)}G_{mn}\nonumber\\
M_{G_{mn},
G_{pq}}&=&-\frac{1}{2\sqrt{G}\exp(-2\P)}(G_{mp}G_{nq}+G_{mq}G_{np})\nonumber\\
M_{B_{mn},
B_{pq}}&=&-\frac{1}{2\sqrt{G}\exp(-2\P)}(G_{mp}G_{nq}-G_{mq}G_{np})\,\,.
\eea
With the help of the functionnal derivatives of the appendix,
one verifies that the two sides are equal. 
This is also the case for all the other components of (\ref{Mt2}).

      \section{Explicit examples}

In this section we study two examples regarding the relationship
between the original effective action and its dual.
The first is a model based on the isometry algebra $SU(2)$.
We consider the sigma model given by following action \cite{bfmps}
\begin{equation}
S=\int d^{2}\sigma \left\{ f(x)\partial _{\mu }x\partial ^{\mu }x
+h(x)Tr\left[ \left( g^{-1}\partial _{\mu }g\right) \left
( g^{-1}\partial ^{\mu }g\right) \right] \right\} 
\end{equation}
where $g$ is an element of the Lie group $SU(2)$ and  $f(x)$ and $ h(x)$
are
arbitrary functions. 
We take the elements of $SU(2)$ to be 
parametrized by the Euler angles $y^i=(u,v,w)$
\begin{equation}
g=\exp (iu\tau_{3})\exp (iv\tau_{1})
\exp (iw\tau _{3})
\end{equation}
where $\tau_i=\sigma_i/2$ with $\sigma_i$ the Pauli matrices.
The only non-vanishing background field is the metric and it is given by 
\begin{equation}
G_{MN}=\left( \begin{array}{cc}
f(x)& 0 \\
 0& h_{ij}(x,y)
\end{array}\right) 
\end{equation}
with 
\begin{equation}
h_{ij}(x,y)=
h(x)\left( \begin{array}{ccc}
1 & 0 & \cos(v)\\
0 & 1 & 0\\
\cos(v) & 0 & 1
\end{array}\right)\,\,. 
\end{equation}
The vielbeins are given by
\begin{equation}
e_{i}^{a}(y)=\left( \begin{array}{ccc}
\sin (v)\sin (w) & \cos (w) & 0\\
-\sin (v)\cos (w) & \sin (w) & 0\\
\cos (v) & 0 & 1
\end{array}\right) 
\end{equation}
where the index $a$ numbers the rows of this matrix.

The objects of the previous sections needed here are 
$S_{ab}(x)=h(x)\delta _{ab}$
and 
$\eta_{ia}f^{a}_{bc}=\varepsilon_{ibc}$
with $\varepsilon_{123}=1$.
These allow the determination of $\bar{S}^{ab}$ and $\bar{A}^{ab}$.
The effective action corresponding to this sigma model is found 
to be given by
\begin{eqnarray}
\Gamma[\omega]  & = & \int dxd^{3}y\sqrt{G}R(G)\nonumber \\
 & = & \frac{3}{2}\int dx du dv dw \left|\sin(v)\right| 
f^{-\frac{3}{2}}h^{\frac{1}{2}}
\left(
f^{2}-2f\frac{d^{2}h}{dx^{2}}+\frac{df}{dx}\frac{dh}{dx}\right)\,\,. 
\label{Ga}
\end{eqnarray}
Notice that one can perform the integration over the internal coordinates
and obtain the internal volume $v=\int du dv dw \left|\sin(v)\right|$. 
The dual background fields are computed from the
duality transformations (\ref{T-rel-simple}).
This yields the following metric, antisymmetric tensor
and dilaton
\begin{eqnarray}
\tilde{G}_{MN}&=&\frac{1}{hD}\left( \begin{array}{cccc}
f & 0 & 0 & 0\\
0 & h^2+u^2 & uv & uw \\
0 & vu & h^2+v^2 & vw\\
0 & wu & wv & h^2+w^2
\end{array}\right) 
\nonumber\\
\tilde{B}_{MN}&=&\frac{1}{D}\left( \begin{array}{cccc}
0 & 0 & 0 & 0 \\
0 & 0 & -w & v \\
0 & w & 0 & -u\\
0 & -v & u & 0
\end{array}\right)
\nonumber\\
 \tilde{\varphi}&=&-\frac{1}{2}\ln(\det M_{ab})=-\frac{1}{2}\ln(hD)\,\,.
\end{eqnarray}
where $D=h^2+u^2+v^2+w^2$.
Using these expressions one can see by direct calculations that 
the T-dual effective action is given by
\begin{eqnarray}
\G[\tilde{\omega}] &=& \int dx d^{3}y \sqrt{\gt}\exp(-2\Pt)
\left\{ \t{R} - \frac{1}{12}\t{H}^2 + 4 \left( 
\t{\nabla}\t{\partial}{\Pt}-(\t{\partial}\Pt)^{2} \right) \right\}
\nonumber\\
&=& \int dx du dv dw 
f^{-\frac{3}{2}}h^{\frac{1}{2}}
\left(
f^{2}-2f\frac{d^{2}h}{dx^{2}}+\frac{df}{dx}\frac{dh}{dx}\right)\,\,.
\label{Gat}
\end{eqnarray}
Similarly, the volume element 
of the internal space in the dual theory 
is $\tilde{v}=\int du dv dw$.
Therefore, the two reduced effective actions $\Gamma[\omega]$ and 
 $\Gamma[\tilde{\omega}]$ are equal up to an overall factor.

Our second example is based on a
Bianchi V type metric. This example was first analysed in \cite{bianchi5}
where the authors found that the $\bb$-functions of the original model 
and its dual are not equivalent. 
The original model is characterized by
\begin{equation}
G_{MN}={\rm diag}
(-f(x),a^{2}(x),a^{2}(x)e^{-2u},a^{2}(x)e^{-2u})\,\,\,,\,\,\, 
B_{MN}=\P=0
\end{equation}  
where the coordinates are such that $z^M=(x,y^i)=(x,u,v,w)$.
The vielbeins of the Bianchi V metric are
\begin{equation}
	e^{a}_{i}={\rm diag}(1,e^{-u},e^{-u})
\end{equation}
and the non-vanishing structure constants 
corresponding to the isometry group are
\begin{equation}
	f_{12}^2 = -f_{21}^2 = f_{13}^3 = -f_{31}^3 = 1 \,\,.
\end{equation}
This example is particularly interesting as $f^{a}_{ab}$ is 
different from zero, while $V_\mu^i$ is zero. 
Therefore equation (\ref{Vef}) is satisfied and we expect
the equality of the reduced effective actions. 
We will see that this is the case here.
However some complications arise at the level of
the equations of motion (the $\bb$-function equations).
By a similar method to the one used in the $SU(2)$ example, 
the dual backgrounds are given by
\begin{eqnarray}
\tilde{G}_{MN}&=&\left( \begin{array}{cccc}
-f & 0 & 0 & 0\\
0 & a^4/D & 0 & 0 \\
0 & 0 & (a^4+w^2)/D & -vw/D\\
0 & 0 & -vw/D & (a^4+v^2)/D
\end{array}\right)
\nonumber\\
\tilde{B}_{MN}&=&\left( \begin{array}{cccc}
0 & 0 & 0 & 0 \\
0 & 0 & -a^2v/D & -a^2w/D \\
0 & a^2v/D & 0 & 0 \\
0 & a^2w/D & 0 & 0
\end{array}\right)
  \nonumber\\ 
\tilde{\varphi}=-\frac{1}{2}\ln(D)
\end{eqnarray}
where $D=a^2(x) \left( a^4(x)+ v^2 + w^2 \right)$.
With these original and dual backgrounds, 
the corresponding effective actions are
\begin{eqnarray}
\G[\omega] &=& 3\int dx du dv dw e^{-2u} 
f^{-\frac{1}{2}}a
\left(  
f\frac{d^{2}a^{2}}{dx^{2}}-a\frac{da}{dx}\frac{df}{dx}-2f^{2}
\right) \nonumber \\
\G[\tilde{\omega}] &=& 3\int dx du dv dw  
f^{-\frac{1}{2}}a
\left(  
f\frac{d^{2}a^{2}}{dx^{2}}-a\frac{da}{dx}\frac{df}{dx}-2f^{2}
\right)\,\,.
\end{eqnarray}

We have then proportionality between the reduced effective actions.
However, this does not mean that the $\bb$-functions are related
according to (\ref{rel_beta}).
Notice that although  $L[\om]=L[\t{\om}]$,
one does not, in general, have $L[\om+\dt \om]=L[\t{\om}+\dt \t{\om}]$.
The difference between $L[\om+\dt \om]$ and $L[\t{\om}+\dt \t{\om}]$
always involves terms containing $f^{a}_{ab}e_{i}^{a}\delta V_{\mu}^i$. 
Clearly, these terms vanish only when $f^a_{ab}=0$.
This is why the beta functions of the original and dual theories
cannot be related by (\ref{rel_beta}) in this case.
This explains the conclusions reached in \cite{bianchi5}
for the particular case
$f(x)=1$ and $a(x)=x$.

\section{Discussion}

We have analysed in this paper the effects of non-Abelian T-duality 
transformations on the Weyl anomaly coefficients of the two-dimensional 
non-linear sigma model. This analysis is carried out at the level 
of the corresponding string effective action.  
A suitable reparametrization of the string backgrounds, in which the
duality
transformations take a simple form, is found. 
In this reparametrization the dual backgrounds are of the same global form 
as the original ones. By an explicit computation, we show that the
Lagrangian 
of the original string effective action is proportional to that of the
dual
string effective action. Using then the connection between the
$\bar\beta$-functions
of the non-linear sigma model and the equations of motion of the string
effective
action, we give a functional relation between the Weyl anomaly
coefficients of the 
original two-dimensional theory and their counterparts in the dual one.  
This is precisely the relation found in \cite{bfmps} through path integral
considerations.

A natural question arises now regarding the invertibility of the
non-Abelian duality transformations.
At the level of the non-linear sigma model, it is well known that these
are not invertible, as 
the dual model does not possess any isometries which allow one to go back 
to the original theory \cite{giveon}.
However, when dealing with the string effective action, 
one can consider the transformations (\ref{Trelations})
in their own right, regardless of their two-dimensional origin. This is
indeed what happens 
in the Abelian case \cite{odd}. Let us therefore examine this important
question.
In order to recover the diagonal part of the metric of the original
theory, namely $h_{ij}$,
from the dual one, that is $\tilde{h}_{ij}$, 
we must impose the following duality transformations
\begin{eqnarray}
{\bar{S}}^{cd} &\rightarrow  & 
\eta ^{ci}e_{i}^{a}S_{ab}e_{j}^{b}\eta ^{jd}\nonumber\\
S_{cd} &\rightarrow  & 
\eta_{ci}E^{i}_{a}\bar{S}^{ab}E^{j}_{b}\eta_{jd}\nonumber\\
e_{i}^{a} & \rightarrow  & \eta_{ib}e_{k}^{b}\eta^{ka}\nonumber\\
{E}^{i}_{a} & \rightarrow  & \eta ^{ib}E^{k}_{b}\eta _{ka}
\label{sbb} 
\end{eqnarray}
This results in the chain of duality transformations 
$h_{ij}\rightarrow \tilde{h}_{ij}=\eta_{ia}\bar{S}^{ab}\eta_{bj}
\rightarrow 
\tilde{h}_{ij}=e_i^aS_{ab}e^b_j$. 
However, these transformations for $S_{ab}$ and $\bar{S}_{ab}$ must
preserve
the relations (\ref{ssbar}). 
This requires that $A_{ab}$ and $\bar{A}_{ab}$ must transform as
\begin{eqnarray}
{\bar{A}}^{cd} &\rightarrow  & 
\eta ^{ci}e_{i}^{a}A_{ab}e_{j}^{b}\eta ^{jd}\nonumber\\
A_{cd} &\rightarrow  & 
\eta_{ci}E^{i}_{a}\bar{A}^{ab}E^{j}_{b}\eta_{jd}
\label{abb} 
\end{eqnarray}  
Using the above duality transformation we automatically get
\begin{eqnarray}
G_{\mu i}&\rightarrow &
 -(G_{\mu k}E_{a}^{k}\bar{A}^{ab}+B_{\mu k}E_{a}^{k}\bar{S}^{ab})\eta
_{bi}
\rightarrow G_{\mu i}\nonumber\\
B_{\mu i}&\rightarrow &
 -(G_{\mu k}E_{a}^{k}\bar{S}^{ab}+B_{\mu k}E_{a}^{k}\bar{A}^{ab})\eta
_{bi}
\rightarrow B_{\mu i}\nonumber\\
G_{\mu \nu}&\rightarrow &
 \tilde{G}_{\mu \nu}
\rightarrow G_{\mu \nu}\nonumber\\
B_{\mu \nu}&\rightarrow &
 \tilde{B}_{\mu \nu}
\rightarrow B_{\mu \nu}
\end{eqnarray}
So far,  all this seems to work quite well but two problems arise:
{}Firstly, the only requirement left for the duality transformation to be
invertible is
\begin{equation}
b_{ij}=e_{i}^{a}v_{ab}e_{j}^{b}\to \eta_{ia}\Ab^{ab}\eta_{bj} \to b_{ij}
\end{equation}
Combining this requirement with the above transformation of
$A_{ab}=v_{ab}+y^{k}\eta _{kc}f_{ab}^{c}$,
implies the strange transformation relation
\begin{equation}
y^{k}\eta _{kc}f_{ab}^{c}e_{i}^{a}e_{j}^{b}
\to 0 \to
y^{k}\eta _{kc}f_{ab}^{c}e_{i}^{a}e_{j}^{b}
\end{equation}
Secondly, using the transformations in (\ref{sbb}), we deduce the
following
chain of transformations for the dilaton 
\begin{equation}
\varphi \rightarrow \varphi +\frac{1}{4}\left(\det \bar{S}
-\det {S}\right) \rightarrow \varphi +\ln \det e
\end{equation}
Therefore, one cannot recover the original dilaton.
It is only in the Abelian case that these two problems do not arise
($f^a_{bc}=0$ and $\det e =1$) and the duality transformations are 
indeed invertible. 

Our results could be of interest in cosmological models
based on the string effective action.
This is due to the fact that most of the relevant models in cosmology, 
exhibit non-Abelian rather than Abelian isometries. Thus, non-Abelian
T-duality transformations could be especially useful in the study
of inflationary models derived from string theories. 
This is in the same spirit as
pre-big-bang inflation \cite{pbb}, where Abelian duality has proven to be 
an essential tool. 

{}Furthermore, our decomposition of the different string backgounds has 
allowed for an explicit computation of the reduced string effective
action.
It is therefore natural to ask if this reduced action presents other
symmetries
like the $O\left({\rm d},{\rm d}\right)$ transformations of the Abelian
case.
This is a crucial point which deserves further study. It is worth
mentioning
that a decomposition, sharing some of the features presented here, has
been used in the 
context of massive supergravity theories \cite{kaloper2}. The authors of
that article 
show that the reduced low enegy effective action of the heterotic string
possesses
an $O\left({\rm d},{\rm d}+16\right)$ symmetry group. However, these
transformations 
present two drawbacks. {}First, the transformations involve the field
strenght of 
the gauge fields coming from the Kaluza-Klein decomposition of the metric. 
This means that these symmetries are non-local on the basic fields.
Secondly, 
the symmetry holds only when a transformation is associated to the
structure constants of the isometry 
Lie algebra.  

There are obviously other subjects to be treated in this area. 
As our analyses are valid only at the one-loop level, it is  therefore
desirable
to examine what happens beyond this leading order. The one-loop duality
transformations must certainly be modified. This could give an idea on how
to correct
the duality symmetry, loop by loop in perturbation theory.  
{}Finally, one could extend the investigation to the case of Poisson-Lie
duality.
This last point is currently under investigation. 
\vskip 2 cm
\noindent
{\bf Aknowledgments:}
We would like to thank Peter Forg\'acs and Ian Jack for useful discussions
and Professor J. M. Lee for his valuable time in explaining the package
Ricci to us.

\newpage
\appendix

\section{Useful Identities}

In section 4 of this paper many quantities 
are redefined for the sake of their simple duality transformations. 
In this appendix, we give the dual 
expressions corresponding to these quantities.
Before doing this, however, one needs  
other useful identities used in the computation.

By applying a  differential operator $\partial$ 
on both sides of the fundamental relations (\ref{ssbar}) one finds
\begin{eqnarray}
\partial \bar{S}^{ab} & = & 
-\partial S_{cd}(S^{ac}S^{db}+A^{ac}A^{db})
-\partial A_{cd}(S^{ac}A^{db}+A^{ac}S^{db})
\nonumber\\
\partial \bar{A}^{ab} & = & 
-\partial S_{cd}(S^{ac}A^{db}+A^{ac}S^{db})
-\partial A_{cd}(S^{ac}S^{db}+A^{ac}A^{db})\,\,.
\label{DSAbar} 
\end{eqnarray} 
When $\partial$ acts on the coordinates $x^\mu$ then 
we have
$\partial _{\mu }A_{cd}= \partial _{\mu }v_{cd}$
while when acting on the coordinates $y^i$ we get
$\partial _{i}S_{cd} = 0 $ 
and 
$\partial _{i}A_{cd}=\eta _{ia}f^{a}_{cd} $.
To identify terms involving $h^{ij}$ and $h_{ij}$ 
in the dual effective action,
we use
\begin{eqnarray}
\label{>h}
e_{i}^{a}(S_{ac}\bar{S}^{cd}S_{db}-A_{ac}\bar{S}^{cd}A_{db})e_{j}^{b}&=&e_{i}^{a}S_{ab}e_{j}^{b}=h_{ij}
\nonumber\\
E_{a}^{i}(\bar{S}^{ab}-\bar{A}^{ac}S_{cd}\bar{A}^{db})E_{b}^{j}&=&E_{a}^{i}S^{ab}E_{b}^{j}=h^{ij}.
\end{eqnarray}
{}Furthermore, the combinations $V_{\mu }^{k}e_{k}^{a}$, $b_{\mu
k}E^{k}_{a}$ and
$b_{\mu k}'E^{k}_{a}$ are independant of $y^i$. Hence, we have the
following identities
\begin{eqnarray}
\partial _{i}V_{\mu }^{j} & = & V_{\mu }^{k}e^{a}_{k}\partial
_{i}E_{a}^{j}\nonumber \\
\partial _{i}b_{\mu j} & = & b_{\mu k}E_{a}^{k}\partial
_{i}e^{a}_{j}\nonumber \\
\partial _{i}b_{\mu j}' & = & b_{\mu k}'E_{a}^{k}\partial
_{i}e^{a}_{j}\,\,.
\label{DVDb} 
\end{eqnarray}
Notice that the Jacobi identity together with the definition of $A_{ab}$
implies
\begin{equation}
A_{cd}f^{d}_{ab}+A_{ad}f^{d}_{bc}+A_{bd}f^{d}_{ca} 
  =  v_{cd}f^{d}_{ab}+{\bf c.p.}
\label{A>v} 
\end{equation}
The above identities lead to the following duality transformations
\begin{eqnarray}
\tilde{V}_{\mu }^{i} & = & (V_{\mu }^{l}e_{l}^{a}y^k\eta_{kc}f^c_{ab}
-b_{\mu l}'E_{b}^{l})\eta ^{bi}\nonumber \\
\t{b}_{\mu i}' & = & -V^{m}_{\mu }e^{a}_{m}\eta _{ai} 
\nonumber \\
\t{b}_{\mu \nu i}' & = & -V^{m}_{\mu \nu }e^{a}_{m}\eta _{ai}
\nonumber \\
\t{b}_{i\mu j}' & = & 0 
\nonumber \\
\t{U}_{\mu \nu } & = & 
-U_{\mu \nu }-2{}F^{k}_{\mu \nu }e^{a}_{k}\lambda _{a} 
\nonumber \\
\t{V}^{i}_{\mu \rho } & = & 
(V_{\mu \rho }^{k}e_{k}^{a}A_{ac}e_m^c
-h^{(1)}_{\mu \rho m})E_{b}^{m}\eta ^{bi} 
\nonumber \\
\t{h}^{(1)}_{\mu \nu i} & = & 
-(V^{k}_{\mu \nu }e^{a}_{k}S_{ac}\bar{S}^{cb}
+h^{(1)}_{\mu \nu k}E^{k}_{a}\bar{A}^{ab})\eta _{bi} 
\nonumber \\
\t{{}F}^{i}_{\mu \nu } & = & 
({}F^{k}_{\mu \nu }e^{a}_{k}A_{ab}e^{b}_{m}
+h^{(2)}_{\mu \nu m})E^{m}_{b}\eta ^{bi} 
\nonumber \\
\t{h}^{(2)}_{\mu \nu i} & = & 
({}F^{k}_{\mu \nu }e^{a}_{k}S_{ac}\bar{S}^{cb}
-h^{(2)}_{\mu \nu k}E^{k}_{a}\bar{A}^{ab})\eta _{bi}\,\,.
\label{derivtrans}
\end{eqnarray}

\section{Comparing the original and dual Lagrangians}

As mentionned in the main body of the paper, 
the terms in $L[\om]$ and $L[\t{\om}]$ are organised
according to the number of factors of $g^{\mu\nu}$ and the number of
factors
of the gauge fields $V_{\mu}^i$, $b_{\mu i}$.
We will gather below these terms according to this criterion.
The expressions coming from the dual Lagrangian $L[\t{\om}]$
are given on the right hand side of each equation.

\subsection{Order zero in $g^{\mu\nu}$ and zero in gauge fields}

To this order one has to collect from the two Lagrangians 
the terms without gauge fields. Many of the terms
coming from the Ricci scalar cancel against terms from 
the dilatonic part (note that this is the case for other orders as well).
The dual expressions of the remainig terms at this order are 
\begin{eqnarray*}
4\t{h}^{ij}\partial_i\partial_j\pst 
-4\t{h}^{ij}\partial _{i}\pst \partial _{j}\pst
+4\partial _{i}(\t{h}^{ij}\partial _{j}\pst )
&=&0\,\,,\\
 -\partial_i \partial _j \t{h}^{ij}&=&
-S^{ab}f^{c}_{ad}f^{d}_{bc}-S^{ab}f^c_{ca}f^d_{db}\,\,,\\
\frac{1}{2} \t{h}^{km} \t{h}^{il} \t{h}^{jn} (\partial_k \t{h}_{ij}
\partial_l \t{h}_{mn}
-\partial_k \t{b}_{ij} \partial_l \t{b}_{mn})&=&
\frac{1}{2}(S^{ab}f^{c}_{ad}f^{d}_{bc}
+S^{ab}A_{cg}S^{gf}A_{eh}S^{hd}f^{c}_{ad}f^{e}_{bf})\,\,,\\
-\frac{1}{4} \t{h}^{kl} \t{h}^{im} \t{h}^{jn} (\partial_{k} \t{h}_{ij}
\partial_{l} \t{h}_{mn}
+\partial_{k} \t{b}_{ij} \partial_{l} \t{b}_{mn})
& = & 
\frac{1}{4}(S_{ab}-A_{ac}S^{cd}A_{db})
(S^{ef}f^{a}_{fg}S^{gh}f^{b}_{he})
\end{eqnarray*}
Similarly, the corresponding expressions at the same order
coming from the original Lagrangian can be easily extracted from the
expressions
given in the main text.  
The sums of these terms in each Lagrangian are identical.
This will be the case at each order as well.
\subsection{Order one in $g^{\mu\nu}$ and zero in gauge fields}
\begin{eqnarray*}
-\frac{1}{4}\t{g}^{\mu \nu }\tilde{h}^{im}\tilde{h}^{jn}
(\partial _{\mu }\tilde{h}_{ij}\partial _{\nu }\tilde{h}_{mn}
+\partial _{\mu }\tilde{b}_{ij}\partial _{\nu }\tilde{b}_{mn})&=&
-\frac{1}{4}g^{\mu \nu }S^{ac}S^{bd}
(\partial _{\mu }S_{ab}\partial _{\nu }S_{cd}
+\partial _{\mu }v_{ab}\partial _{\nu }v_{cd})\\
&=&-\frac{1}{4}g^{\mu \nu }h^{im}h^{jn}
(\partial _{\mu }h_{ij}\partial _{\nu }h_{mn}
+\partial _{\mu }b_{ij}\partial _{\nu }b_{nm})
\end{eqnarray*}
\subsection{Order one in $g^{\mu\nu}$ and one in gauge fields}
To this order one must demand that
\be
\partial _{k}V^{\alpha k}-2V^{\alpha k}\partial _{k}\psi =0
\ee
in order to get rid of the terms involving the field $\psi$. 
Note that this is equivalent to imposing equation (\ref{Vef}).
The duals of the remaining terms at this order are  
\bea
\t{g}^{\alpha\beta}\partial_{i}\t{V}_{\beta}^j\partial_{\alpha
}\t{h}_{jk}\t{h}^{ki}
&+&\t{g}^{\alpha \beta }\t{h}^{im}\t{h}^{jn}\partial_{\alpha }
\t{h}_{ij}\partial_{m}\t{V}^{k}_{\beta }\t{b}_{kn}\nonumber\\
&=& g^{\alpha\beta}V_{\alpha}^ie_i^af^b_{ac}
(S^{cd}A_{be}S^{ef}\partial_{\beta}A_{fd}-S^{cd}\partial_{\beta}S_{db})\,\,,
\nonumber\\
\frac{1}{2}\t{V}^{\alpha k}\t{h}^{im}\t{h}^{jn}\partial_{\alpha
}\t{h}_{ij}
\partial_{k}\t{h}_{mn}
&+&\frac{1}{2}\t{g}^{\alpha \beta }\t{h}^{im}\t{h}^{jn}
\partial_{\alpha }\t{h}_{ij}\t{V}^{k}_{\beta }\partial_{k}\t{b}_{mn}
\nonumber\\
&=&  \frac{1}{2}g^{\mu \nu }\tilde{V}^{i}_{\mu }S^{ab}S^{cd}
\partial _{\nu }A_{ac}\partial _{i}A_{bd}\,\,,\nonumber\\
\t{g}^{\alpha \beta }\t{h}^{im}\t{h}^{jn}\partial_{\alpha }
\t{h}_{ij}\partial_{m}\t{b}_{\beta n}'& = & 0
\eea
\subsection{Order one in $g^{\mu\nu}$ and two in gauge fields}
\begin{eqnarray}
-\frac{1}{2}\t{g}^{\mu \nu }\partial _{i}\t{V}^{j}_{\mu }\partial
_{j}\t{V}^{i}_{\nu } 
& = & 
-\frac{1}{2}g^{\mu \nu }V^{k}_{\mu }e_{k}^{a}V^{l}_{\nu
}e_{l}^{b}f^{c}_{ad}f^{d}_{bc}
\nonumber\\
-\frac{1}{2}\t{g}^{\mu \nu }
\t{h}^{ik}\t{h}_{jl}\partial _{i}\t{V}^{j}_{\mu }
\partial _{k}\t{V}^{l}_{\nu } 
&-&\frac{1}{2}\t{g}^{\mu \nu }\t{h}^{im}\t{h}^{jn}\partial _{i}
\t{V}^{l}_{\mu }\t{b}_{lj}(\partial _{m}\t{V}^{k}_{\nu
}\t{b}_{kn}-(m\leftrightarrow n))\nonumber\\
& = & 
-\frac{1}{2}g^{\mu \nu }V^{k}_{\mu }e_{k}^{a}
V^{l}_{\nu }e_{l}^{b}f^{c}_{ae}f^{d}_{bf}(S_{cd}-A_{cg}S^{gh}A_{hd})S^{ef}
\nonumber \\
 & & 
+\frac{1}{2}g^{\mu \nu }V^{k}_{\mu }e_{k}^{a}
V^{l}_{\nu }e_{l}^{b}f^{c}_{ae}f^{d}_{bf}A_{cg}S^{gf}A_{dh}S^{he}\,\,,
\nonumber\\
-\t{g}^{\mu \nu }\t{h}^{il}\t{V}_{\mu }^{k}\partial _{k}\t{h}_{lj}\partial
_{i}\t{V}^{j}_{\nu }
&-&\t{g}^{\mu \nu }\t{h}^{im}\t{h}^{jn}\t{V}^{l}_{\mu }
\partial _{l}\t{b}_{ij}\partial _{m}\t{V}^{k}_{\nu }\t{b}_{kn}\nonumber\\
& = & 
g^{\mu \nu }S^{cd}S^{ef}(b_{\mu k}'E^{k}_{a}V^{l}_{\nu
}e_{l}^{b}f_{ce}^{a}f^{g}_{bf}A_{gd}
\nonumber\\
&&+V^{k}_{\mu }e^{a}_{k}V^{l}_{\nu
}e^{b}_{l}y^i\eta_{ij}f^j_{ag}f_{de}^{g}A_{ch}f_{bf}^{h})\,\,,
\nonumber\\
-\frac{1}{4}\t{V}^{i}_{\mu }\t{V}^{j}_{\nu }\t{h}^{mp}h^{nq}
\partial _{i}\t{h}_{mn}\partial _{j}\t{h}_{pq}
&-&\frac{1}{4}\t{g}^{\mu \nu }\t{h}^{im}\t{h}^{jn}\t{V}^{k}_{\mu
}\t{V}^{l}_{\nu }
\partial _{k}\t{b}_{ij}\partial _{l}\t{b}_{mn}\nonumber\\
& = & 
-\frac{1}{4}g^{\mu \nu }S^{ac}S^{bd}b_{\mu i}'E^{i}_{e}f_{ab}^{e}b_{\nu
j}'E^{j}_{f}f_{cd}^{f}
\nonumber \\
 &  &-g^{\mu \nu }S^{cd}S^{ef}b_{\mu k}'E^{k}_{a}V^{l}_{\nu
}e_{l}^{b}f_{ce}^{a}f^{g}_{bf}
y^i\eta_{ij}f^j_{gd}
\nonumber \\
 &  & -\frac{1}{4}g^{\mu \nu }S^{cd}S^{ef}V^{k}_{\mu }e_{k}^{a}
V^{l}_{\nu
}e^{b}_{l}y^p\eta_{pq}f^q_{ag}f_{ce}^{g}y^i\eta_{ij}f^j_{bh}f_{df}^{h}\,\,,
\nonumber\\
\t{g}^{\mu \nu }\t{h}^{im}\t{h}^{jn}\t{V}^{l}_{\mu }\partial
_{l}\t{b}_{mn}
\partial _{i}\t{b}_{\nu j}' & = & 0
\nonumber\\
\t{g}^{\mu \nu }\t{h}^{im}\t{h}^{jn}\partial _{m}\t{V}^{k}_{\nu
}\t{b}_{kn}
(\partial _{i}\t{b}_{\nu j}'-(i\leftrightarrow j)) & = & 0
\nonumber\\
-\frac{1}{2}\t{g}^{\mu \nu }\t{h}^{im}\t{h}^{jn}\partial _{i}\t{b}_{\mu
j}'
(\partial _{m}\t{b}_{\nu n}'-(m\leftrightarrow n)) & = & 0
\end{eqnarray}

\subsection{Order two in $g^{\mu\nu}$}

Here we can treat at the same time order two, three and four 
in the gauge fields.
Using the transformations (\ref{derivtrans}) 
and the relations (\ref{>h})
we can easily show that
 \begin{eqnarray*}
\tilde{h}_{ij}
(\tilde{V}^{i}_{\mu\nu}-\tilde{{}F}^{i}_{\mu\nu})
(\tilde{V}^{j}_{\lambda\sigma}-\tilde{{}F}^{j}_{\lambda\sigma})
&+&
\tilde{h}^{ij}
(\tilde{h}_{\mu \nu i}^{(1)}+\tilde{h}_{\mu \nu i}^{(2)})
(\tilde{h}_{\lambda\sigma j}^{(1)}+\tilde{h}_{\lambda\sigma j}^{(2)})
\\ 
& = & 
(V^{i}_{\mu \nu }-{}F^{i}_{\mu \nu })
e_{i}^{a}(S_{ac}\bar{S}^{cd}S_{db}-A_{ac}\bar{S}^{cd}A_{db})e_{j}^{b}
(V^{j}_{\lambda \sigma }-{}F^{j}_{\lambda \sigma })
\\
&& 
+(h_{\mu \nu i}^{(1)}+h_{\mu \nu i}^{(2)})
E_{a}^{i}(\bar{S}^{ab}-\bar{A}^{ac}S_{cd}\bar{A}^{db})E_{b}^{j}
(h_{\lambda \sigma j}^{(1)}+h_{\lambda \sigma j}^{(2)}) 
\\
&=&
h_{ij}
(V^{i}_{\mu \nu }-{}F^{i}_{\mu \nu })
(V^{j}_{\lambda \sigma }-{}F^{j}_{\lambda \sigma })
+
h^{ij}
(h_{\mu \nu i}^{(1)}+h_{\mu \nu i}^{(2)})
(h_{\lambda \sigma j}^{(1)}+h_{\lambda \sigma j}^{(2)}) 
\end{eqnarray*}

\subsection{Order three in $g^{\mu\nu}$: 
the invariance of $ h_{\mu \nu \rho }$}

Starting from the definition of $ h_{\mu \nu \rho }$ 
given in (\ref{h_MNP}) and using the decomposition 
(\ref{B_MN}) one can show that
\be
h_{\mu \nu \rho }=
\partial_{\rho}b_{\mu\nu}-V_{\rho}^k\partial_k b_{\mu\nu}
+\frac{1}{2}(V^{k}_{\mu \rho }b_{\nu k}'
+b_{\mu \rho k}'V_{\nu }^{k})
+ \frac{1}{2}({}F^{k}_{\rho \mu }b_{\nu l}'
+V_{\rho }^{k}V_{\mu }^{l}b_{l\nu k}')
+{\bf c.p.}
\label{hdec}
\ee
Now using the transformations (\ref{derivtrans}) 
and the key relations (\ref{ssbar})
we find that 
\begin{eqnarray*}
\frac{1}{2}(\tilde{V}^{k}_{\mu \rho }\tilde{b}_{\nu k}'
+\tilde{b}_{\mu \rho k}'\tilde{V}_{\nu }^{k})+{\bf c.p.}
&=&
\frac{1}{2}(V^{k}_{\mu \rho }b_{\nu k}'
+b_{\mu \rho k}'V_{\nu }^{k})+{\bf c.p.}
\\
\frac{1}{2}(\t{{}F}^{k}_{\rho \mu }\t{b}_{\nu k}'
+\t{V}_{\rho }^{k}\t{V}_{\mu }^{l}\t{b}_{l\nu k}')+{\bf c.p.}
&=&
\frac{1}{2}({}F^{k}_{\rho \mu }b_{\nu k}'
+V_{\rho }^{k}V_{\mu }^{l}b_{l\nu k}')+{\bf c.p.}
\end{eqnarray*}
Moreover, the term $b_{\mu\nu}$ is invariant and independant of $y^i$.
All this applied to the decomposition (\ref{hdec})
proves the invariance of $ h_{\mu \nu \rho }$ under non-Abelian T-duality.

\section{{}Functional derivatives}

Here, we list 
the non-vanishing components of $\frac{\dt \t{\om}_A}{\dt \om_B}$. 
These are needed to prove equation (\ref{Mt2}).
With the two definitions
\bea
Y_{\mu}^n
&=&(G_{\mu i} E^i_a \bar{S}^{ab}+B_{\mu i} E^i_a
\bar{A}^{ab})E_b^n\nonumber\\
Z_{\mu}^n
&=&(B_{\mu i} E^i_a \bar{S}^{ab}+G_{\mu i} E^i_a \bar{A}^{ab})E_b^n
\eea
we have 
\bea
\frac{\dt \t{G}_{\mu\nu}}{\dt
{G}_{\lambda\sigma}}&=&\frac{1}{2}[\dt_{\mu}^{\lambda}\dt_{\nu}^{\sigma}
+(\mu \leftrightarrow \nu)] 
\nonumber\\
\frac{\dt \t{G}_{\mu\nu}}{\dt {G}_{\lambda n}}&=&
-[Y_{\mu}^n\dt_{\nu}^{\lambda}
+(\mu \leftrightarrow \nu)]\nonumber\\
\frac{\dt \t{G}_{\mu\nu}}{\dt {B}_{\lambda n}}&=&
[Z_{\mu}^n\dt_{\nu}^{\lambda}
+(\mu \leftrightarrow \nu)]\nonumber\\
\frac{\dt \t{G}_{\mu\nu}}{\dt {G}_{m n}}&=&
\frac{1}{2}\{
Y_{\mu}^mY_{\nu}^n-Z_{\mu}^mZ_{\nu}^n
+(\mu \leftrightarrow \nu)\}
\nonumber\\
\frac{\dt \t{G}_{\mu\nu}}{\dt {B}_{m n}}&=&
\frac{1}{2}\{Z_{\mu}^mY_{\nu}^n-Y_{\mu}^mZ_{\nu}^n
-(\mu \leftrightarrow \nu)\}
\nonumber\\
\frac{\dt \t{G}_{\mu i}}{\dt {G}_{\lambda
n}}&=&-\dt_{\mu}^{\lambda}E^n_a\bar{A}^{ab}\eta_{bi}
\nonumber\\
					       &=&\frac{\dt \bar{B}_{\mu
i}}{\dt {B}_{\lambda n}}
\nonumber\\	
\frac{\dt \t{G}_{\mu i}}{\dt {B}_{\lambda
n}}&=&-\dt_{\mu}^{\lambda}E^n_a\bar{S}^{ab}\eta_{bi}
\nonumber\\
					       &=&\frac{\dt \bar{B}_{\mu
i}}{\dt {G}_{\lambda n}}
\nonumber\\	
\frac{\dt \t{G}_{\mu i}}{\dt {G}_{m n}}&=&
\frac{1}{2}\{Z_{\mu}^mE^n_c\bar{S}^{cd}\eta_{di}+Y_{\mu}^mE^n_c\bar{A}^{cd}\eta_{di}
+(m \leftrightarrow n)\}
\nonumber\\	
\frac{\dt \t{G}_{\mu i}}{\dt {B}_{m n}}&=&
\frac{1}{2}\{Y_{\mu}^mE^n_c\bar{S}^{cd}\eta_{di}+Z_{\mu}^mE^n_c\bar{A}^{cd}\eta_{di}
-(m \leftrightarrow n)\}
\nonumber\\	
\frac{\dt \t{G}_{ij}}{\dt {G}_{m n}}&=& 
-\frac{1}{2}\{\eta_{ia}[\bar{S}^{ab}E_b^mE^n_c\bar{S}^{cd}+\bar{A}^{ab}E_b^mE^n_c\bar{A}^{cd}]\eta_{dj}
+(m \leftrightarrow n)\}
\nonumber\\	
\frac{\dt \t{G}_{ij}}{\dt {B}_{m n}}&=& 
-\frac{1}{2}\{\eta_{ia}[\bar{S}^{ab}E_b^mE^n_c\bar{A}^{cd}+\bar{A}^{ab}E_b^mE^n_c\bar{S}^{cd}]\eta_{dj}
-(m \leftrightarrow n)\}
\nonumber\\	
\frac{\dt \Pt}{\dt {B}_{m n}}&=& -\frac{1}{2}E^m_a\bar{S}^{ab}E_b^n
\nonumber\\	
\frac{\dt \t{B}_{\mu\nu}}{\dt
{B}_{\lambda\sigma}}&=&\frac{1}{2}[\dt_{\mu}^{\lambda}\dt_{\nu}^{\sigma}
-(\mu \leftrightarrow \nu)] 
\nonumber\\
\frac{\dt \t{B}_{\mu\nu}}{\dt {G}_{\lambda n}}&=&
-[Z_{\mu}^n\dt_{\nu}^{\lambda}
-(\mu \leftrightarrow \nu)]\nonumber\\
\frac{\dt \t{B}_{\mu\nu}}{\dt {B}_{\lambda n}}&=&
[Y_{\mu}^n\dt_{\nu}^{\lambda}
-(\mu \leftrightarrow \nu)]\nonumber\\
\frac{\dt \t{B}_{\mu\nu}}{\dt {B}_{m n}}&=&
\frac{1}{2}\{
Y_{\mu}^mY_{\nu}^n-Z_{\mu}^mZ_{\nu}^n
-(\mu \leftrightarrow \nu)\}
\nonumber\\
\frac{\dt \t{B}_{\mu\nu}}{\dt {G}_{m n}}&=&
\frac{1}{2}\{Z_{\mu}^mY_{\nu}^n-Y_{\mu}^mZ_{\nu}^n
-(\mu \leftrightarrow \nu)\}
\nonumber\\
\frac{\dt \t{B}_{\mu i}}{\dt {B}_{m n}}&=&
\frac{1}{2}\{Z_{\mu}^mE^n_c\bar{S}^{cd}\eta_{di}+Y_{\mu}^mE^n_c\bar{A}^{cd}\eta_{di}
-(m \leftrightarrow n)\}
\nonumber\\	
\frac{\dt \t{B}_{\mu i}}{\dt {G}_{m n}}&=&
\frac{1}{2}\{Y_{\mu}^mE^n_c\bar{S}^{cd}\eta_{di}+Z_{\mu}^mE^n_c\bar{A}^{cd}\eta_{di}
+(m \leftrightarrow n)\}
\nonumber\\	
\frac{\dt \t{B}_{ij}}{\dt {B}_{m n}}&=& 
-\frac{1}{2}\{\eta_{ia}[\bar{S}^{ab}E_b^mE^n_c\bar{S}^{cd}+\bar{A}^{ab}E_b^mE^n_c\bar{A}^{cd}]\eta_{dj}
-(m \leftrightarrow n)\}
\nonumber\\	
\frac{\dt \t{B}_{ij}}{\dt {G}_{m n}}&=& 
-\frac{1}{2}\{\eta_{ia}[\bar{S}^{ab}E_b^mE^n_c\bar{A}^{cd}+\bar{A}^{ab}E_b^mE^n_c\bar{S}^{cd}]\eta_{dj}
+(m \leftrightarrow n)\}
\nonumber\\	
\frac{\dt \Pt}{\dt {B}_{m n}}&=& \frac{1}{2}E^m_a\bar{A}^{ab}E_b^n
\nonumber\\	
\frac{\dt \Pt}{\dt \P}&=& 1\,\,.
\label{funcderiv}
\eea

\end{document}